\documentclass[11pt]{article}
\usepackage{amsmath}
\usepackage{graphicx}
\usepackage{amsfonts}
\usepackage{graphics}
\usepackage{natbib}

\newcommand{\argmin}{\operatornamewithlimits{argmin}}

\newcommand{\sRCV}{\mbox{\scriptsize RCV}}
\newcommand{\wRCV}{\mbox{\scriptsize WRCV}}

\newcommand{\sLL}{\mbox{\scriptsize LL}}
\newcommand{\sCVL}{\mbox{\scriptsize CVL}}
\newcommand{\sSCAD}{\mbox{\scriptsize SCAD}}
\newcommand{\E}{\mbox{E}}
\newcommand{\Var}{\mbox{Var}}
\newcommand{\bs}{\boldsymbol}
\newcommand{\beps}{\boldsymbol{\varepsilon}}

\newcommand{\bx}{\mbox{\bf x}}

\newcommand{\bB}{\mbox{\bf B}}

\newcommand{\bS}{\mbox{\bf S}}

\newcommand{\bX}{\mbox{\bf X}}

\newcommand{\bY}{\mbox{\bf Y}}

\newcommand{\bbeta}{\mbox{\boldmath $\beta$}}

\newcommand{\var}{\mathrm{var}}
\newcommand{\cov}{\mathrm{cov}}
\newcommand{\corr}{\mathrm{corr}}

\newtheorem{theorem}{Theorem}

\newtheorem{lemma}{Lemma}

\def\beps{{ \boldsymbol{\varepsilon}}}
\def\pp{{\parallel}}
\begin{document}
\title{Variance Estimation Using Refitted Cross-validation in Ultrahigh Dimensional Regression}
\author{Jianqing Fan, Shaojun Guo and Ning Hao}

\date{}
\maketitle
\begin{abstract}
Variance estimation is a fundamental problem in statistical modeling. In ultrahigh dimensional linear regression where the dimensionality
is much larger than sample size, traditional variance estimation techniques are not applicable. Recent advances on variable selection in ultrahigh dimensional linear regression make this problem accessible. One of the major problems in ultrahigh dimensional regression is the high spurious correlation between the unobserved realized noise and some of the predictors.  As a result, the realized noises are actually predicted when extra irrelevant variables are selected, leading to serious underestimate of the noise level. In this paper, we propose a two-stage refitted procedure via a data splitting technique, called refitted cross-validation (RCV), to attenuate the influence of irrelevant variables with high spurious correlations. Our asymptotic results show that the resulting procedure performs as well as the oracle estimator, which knows in advance the mean regression function. The simulation studies lend further support to our theoretical claims.  The naive two-stage estimator and the plug-in one stage estimators using LASSO and SCAD are also studied and compared. Their performances can be improved by the proposed RCV method.
\end{abstract}
%\keywords{Data splitting; Dimension reduction; High dimensionality; Refitted cross-validation; Sure Screening; Variance estimation; Variable selection.}

%%%%%%%%%%%%%%%%%%%%%%%%%%%%%%%%%%%%%%%%%%%%%%%%%%%%%%%%%%%%%%%%%%%%%%%%%%%%%%%%%%%%%%%%%%%
%
%                Address for Correspondence
%
%%%%%%%%%%%%%%%%%%%%%%%%%%%%%%%%%%%%%%%%%%%%%%%%%%%%%%%%%%%%%%%%%%%%%%%%%%%%%%%%%%%%%%%%%%%
%\footnotetext{{\it Address for Correspondence}: Jianqing Fan, Department of Operational Research $\&$ Financial
%Engineering, Princeton University, Princeton,
%New Jersey 08544, USA.  The paper was supported by the NIH Grant R01-GM072611 and NSF Grants DMS-0704337 and was completed while Shaojun Guo and Ning Hao were a postdoctoral fellow at Princeton University.\\
%Email: jqfan@princeton.edu.\\
%{\bf Dedicated to Peter J. Bickel on his occasion of the 70th birthday.}}

\footnotetext[1]{Jianqing Fan is Frederick L. Moore Professor of Finance, Department of Operational Research $\&$ Financial Engineering, Princeton University, Princeton, NJ 08544, U.S.A. (Email: jqfan@princeton.edu). Shaojun Guo is Assistant Professor, Institute of Applied Mathematics, Academy of Mathematics and Systems Science, Chinese Academy of Sciences, Beijing 100190, P.R.China (E-mail: guoshaoj@amss.ac.cn). Ning Hao is Visiting Assistant Professor, Department of Mathematics, the University of Arizona, Tucson, AZ 85721, U.S.A. (Email: nhao@math.arizona.edu). The paper was supported by the NIH Grant R01-GM072611 and NSF Grants DMS-0704337 and was completed while Shaojun Guo and Ning Hao were a postdoctoral fellow at Princeton University. \\
{\bf Dedicated to Peter J. Bickel on his occasion of the 70th birthday.}
}
\large
\section{Introduction}

Variance estimation is a fundamental problem in statistical modeling.  It is prominently featured in the statistical inference on regression coefficients. It is also important for variable selection criteria such as AIC and BIC. It provides also a benchmark of forecasting error when an oracle actually knows the regression function and such a benchmark is very important for forecasters to gauge their forecasting performance relative to the oracle. For conventional linear models, the residual variance estimator usually performs well and plays an important role in the inferences after model selection and estimation. However, the ordinary least squares methods don't work for many contemporary datasets which have more number of covariates than the sample size. For example, in disease classification using microarray data, the number of arrays is usually in tens, yet tens of thousands of gene expressions are potential predictors. When interactions are considered, the dimensionality grows even more quickly, e.g. considering possible interactions among thousands of genes or SNPs yields the number of parameters in the order of millions. In this paper, we propose and compare several methods for variance estimation in the setting of ultrahigh dimensional linear model.
%Throughout the paper, high dimensionality refers to the case where the number of covariates grows at polynomial rate of the sample size, and ultrahigh dimensionality refers to the case where the number of covariates grows at non-polynomial rate.
A key assumption which makes the high dimensional problems solvable is the sparsity condition: the number of nonzero components is small compared to the sample size. With sparsity, variable selection can identify the subset of important predictors and improve the model interpretability and predicability.

Recently, there have been several important advances in model selection and estimation for ultrahigh dimensional problems. The properties of penalized likelihood methods such as the LASSO and SCAD have been extensively studied in high and ultrahigh dimensional regression.  Various useful results have been obtained. See, for example, \cite{ISI:000221981400004,ISI:000245390700010,bune:tsyb:spar:2007, ISI:000258243000006,mein:yu:lass:2009, Kim:Choi:OH:SCADHD:2008, MeierGB08, LvFan09, Fan:Lv:NP:2009}. Another important model selection tool is the Dantzig selector proposed by \citet{CandesTao07} which can be easily recast as a linear program. It is closely related to LASSO, as demonstrated by \cite{bick:rito:simu:2008}. \citet{FanLv08} showed that correlation ranking possesses a sure screening property in the Gaussian linear model with Gaussian covariates and proposed a sure independent screening (SIS) and iteratively sure independent screening (ISIS) method. \citet{Fan:Samworth:Wu:2008} extended ISIS to a general pseudo-likelihood framework, which includes generalized linear models as a special case. \citet{FanSong2010} have developed general conditions under which the marginal regression possesses a sure screening property in the context of generalized linear model. For an overview, see \citet{FanLv10}.

In all the work mentioned above, the primary focus is the consistency of model selection and parameter estimation. The problem of variance estimation in ultrahigh dimensional setting has hardly been touched. A natural approach to estimate the variance is the following two-stage procedure. In the first stage, a model selection tool is applied to select a model which, if is not exactly the true model, includes all important variables with moderate model size (smaller than the sample size). In the terminology of \cite{FanLv08}, the selected model has a sure screening property.
In the second stage, the variance is estimated by ordinary least squares method based on the selected variables in the first stage. Obviously, this method works well if we are able to recover exactly the true model in the first stage. This is usually hard to achieve in ultrahigh dimensional problems. Yet, sure screening properties are much easier to obtain. Unfortunately, this naive two-step approach can seriously underestimate the noise level even with the sure screening property in the first stage due to spurious correlation inherent in ultrahigh dimensional problems. When the number of irrelevant variables is huge, some of these variables have large sample correlations with the realized noises. Hence, almost all variable selection procedures will, with high probability, select those spurious variables in the model when the model is over fitted, and the realized noises are actually predicted by several spurious variables, leading to serious underestimate of the residual variance.

\begin{figure}  %Figure 1
\includegraphics[width = 162mm, height = 80mm]{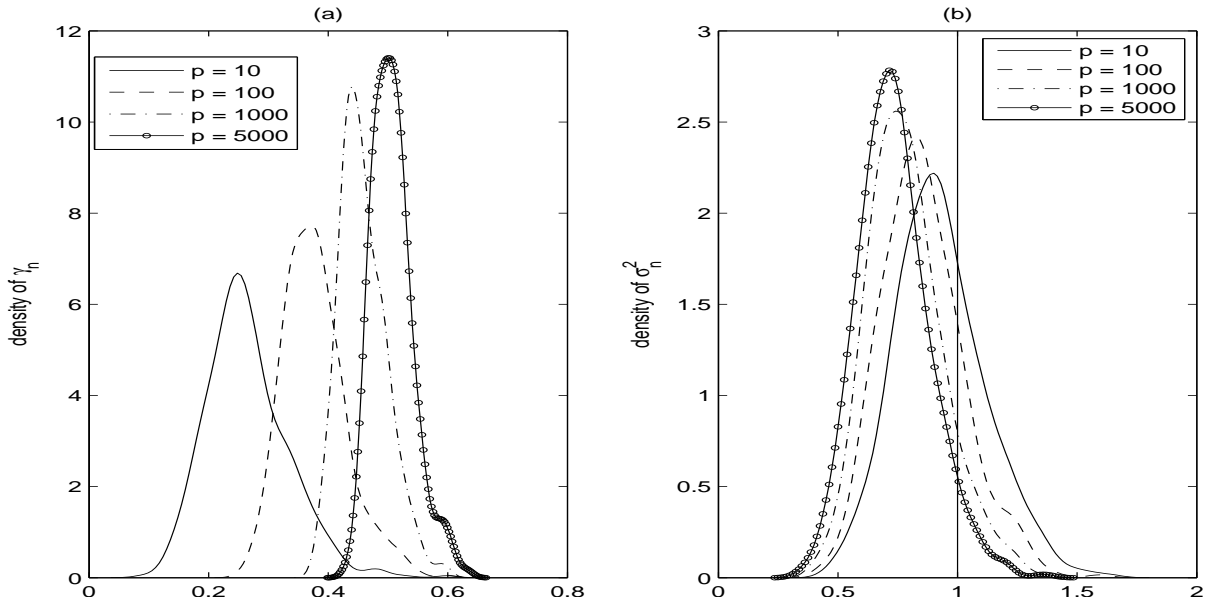}
\caption{(a) Densities of the maximum absolute sample correlation $\gamma_n$ for various $p$. (b)
Densities of the corresponding estimates $\hat{\sigma}_n^2$ given by (\ref{a1}).  The vertical line marks the true variance 1.  All calculations are based on 500 simulations and the sample size $n$ is $ 50$.} \label{fig1}
\end{figure}

The above phenomenon can be easily illustrated in the simplest model, in which the true coefficient $\bbeta=0$. %This does not lose the generality, as the sure screening property ensures the non-vanishing parameters are estimated consistently by the relevant variables and the remaining variables are selected to predict the realized noise vector.
Suppose that one extra variable is selected by a method such as the LASSO or SIS in the first stage. Then, the ordinary least squares estimator $\hat{\sigma}_n^2$ is
\begin{eqnarray} \label{a1}
\hat{\sigma}_n^2=(1-\gamma_n^2){1\over n-1}\sum_{i=1}^{n}(Y_i-\bar{Y})^2.
\end{eqnarray}
where $\gamma_n$ is the sample correlation of the spurious variable and the response, which is really the realized noise in this null model.  Most variable selection procedures such as stepwise addition, SIS and LASSO will first select the covariate that has highest sample correlation with the response, namely, $\gamma_n=\max_{j \leq p}|\widehat{\corr}_n(X_j,Y)|$.  In other words, this extra variable is selected to best predict the realized noise vector. However, as \cite{FanLv08} stated, the maximum absolute sample correlation $\gamma_n$ can be very large, which makes  $\hat{\sigma}_n^2$ seriously biased. To illustrate the point, we simulated 500 data sets with sample size $n=50$ and the number of covariates $p=10$, $100$, $1000$ and $5000$, with $\{X_j\}_{j=1}^p$ and noise $i.i.d.$ from the standard normal distribution. Figure~\ref{fig1}(a) presents the densities of $\gamma_n$ across the 500 simulations and Figure~\ref{fig1}(b) depicts the densities of the estimator $\hat{\sigma}_n^2$ defined in (\ref{a1}). Clearly, the biases of $\hat{\sigma}_n^2$ become larger as $p$ increases.

The bias gets larger when more spurious variables are recruited in the model.  To illustrate the point, let us use the stepwise addition to recruit $s$ variables to the model.  Clearly, the realized noises are now better predicted, leading to even more severe underestimate of the noise level.  Figure 2 depicts the distributions of spurious multiple correlation with the response (realized noise) and the corresponding naive two-stage estimator of variance for $s=1, 2, 5$ and $10$, keeping $p = 1000$ fixed.  Clearly, the biases get much larger with $s$.  For comparison, we also depict similar distributions based on SIS, which selects $s$ variables that are marginally most correlated with the response variable.  The results are depicted in Figure 3 (a).  While the biases based on the SIS method are still large, they are smaller than those based on the stepwise addition method, as the latter chose the coordinated spurious variables to optimize the prediction of the realized noise.

A similar phenomenon was also observed in classical model selection by \citet{Ye:on:1998}. To correct the effects of model selection, \citet{Ye:on:1998} developed a concept of generalized degree of freedom (GDF) but it is computationally intensive and can only be applied to some special cases.
%It is unclear whether his method can be effectively extended to the high or ultrahigh dimensional settings.

 \vspace{3ex}
\begin{figure}
\includegraphics[width = 162mm, height = 80mm]{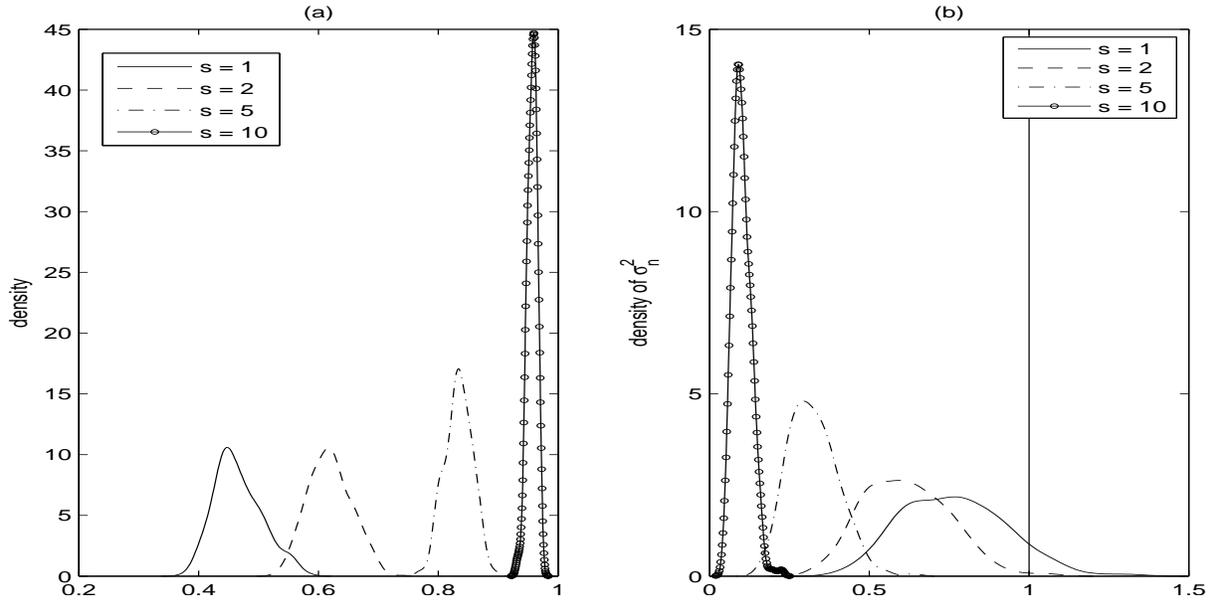}
\caption{(a) Densities of spurious multiple correlation with the response for various number of spurious variables $s$.
 (b) Densities of the naive two-stage estimators of variance.  All calculations are based on stepwise addition algorithm with 500 simulations, $n =50$, and $p = 1000$.
 The vertical line marks the true variance 1. } \label{fig2}
\end{figure}

\begin{figure}
\includegraphics[width = 162mm, height = 80mm]{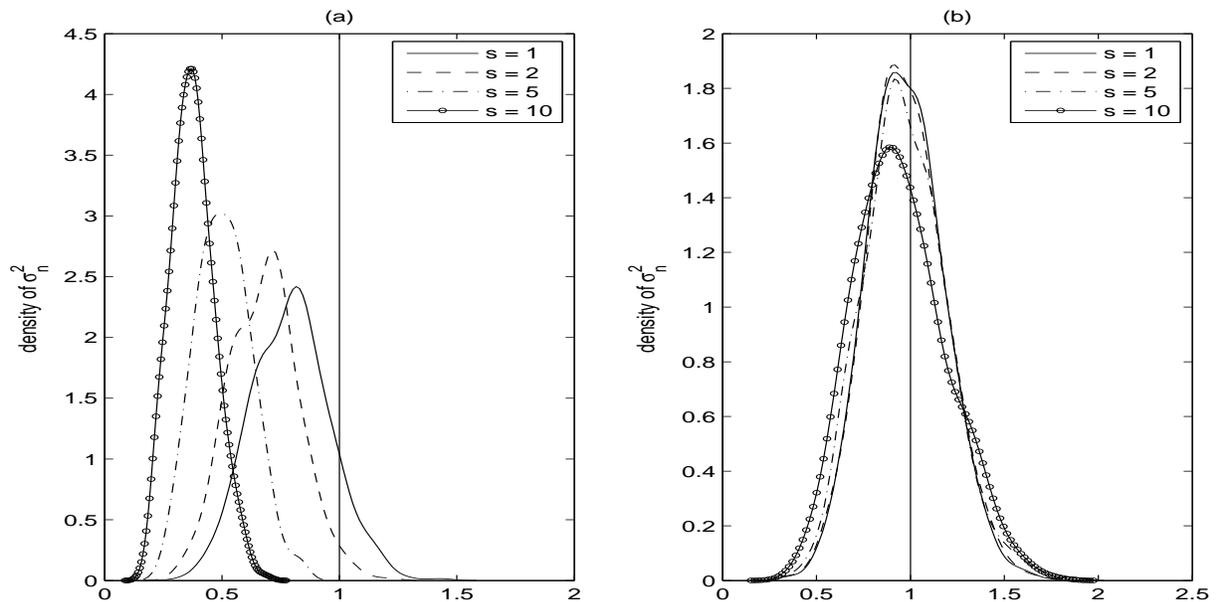}
\caption{(a) Densities of the variance estimators based on the naive two-stage approach with number of spurious variables $s=1, 2, 5$, and 10.
 (b) Densities of RCV estimators of variance.  All calculations are based on 500 simulations using SIS as a model selector and the sample size $n$ is $50$. They show that the biases of the naive two-stage estimator are correctable. The vertical line marks the true variance 1. } \label{fig3}
\end{figure}

To attenuate the influence of spurious variables entered into the selected model and to improve the estimation accuracy, we introduce a refitted cross-validation (RCV) technique. Roughly speaking, we split the data randomly into two halves, do model selection using the first half dataset, and refit the model based on the variables selected in the first stage, using the second half data, to estimate the variance, and vice versa. The proposed estimator is just the average of these two estimators. The results of the RCV variance estimators with $s=1, 2, 5$ and 10 are presented in Figure 3(b).  The corrections of biases due to spurious correlation are dramatic. The essential difference of this approach and the naive two-stage approach is that the regression coefficients in the first stage are discarded and refitted using the second half data and hence the spurious correlations in the first stage are significantly reduced at the second stage. The variance estimation is unbiased as long as the selected models in the first stage contain all relevant variables, namely, possess a sure screening property.  It turns out that this simple RCV method improves dramatically the performance of the naive two-stage procedure.  Clearly, the RCV can also be used to do model selection itself, reducing the influence of spurious variables.

%To appreciate why, suppose a predictor has a big sample correlation with the response (realized noise in the null model) over the first half dataset and is selected into the model by a model selection procedure. The independence of the two datasets imply that it is very unlikely that this variable has a large sample correlation with the response over the second half of the dataset. Hence, it will hardly affect the estimation of the variance when model is refitted for the second dataset.

To appreciate why, suppose a predictor has a big sample correlation with the response (realized noise in the null model) over the first half dataset and is selected into the model by a model selection procedure. Since the two halves of the dataset are independent and the chance that a given predictor is highly correlated with realized noise is small, it is very unlikely that this predictor has a large sample correlation with the realized noise over the second half of the dataset.  Hence, its impact on the variance estimation is very small when refitted and
%the selected predictor is nothing special with respect to the second half.
estimating the variance over the second half will not cause any bias. The above argument is also true for the non-null models provided that the selected model includes all important variables.

%it should be not predictive whether or not this variable has a large sample correlation with the response over the second half of the dataset, just as in a coin tossing experiment, the outcome of the first flip does not predict the outcome of the second flip. Hence, this variable will hardly affect the estimation of the variance when model is refitted for the second dataset.

To gain better understanding of the RCV approach, we compare our method with the direct plug-in method, which computes the residual variance based on a regularized fit. This is inspired by \citet{Greenshtein:Ritov:persistence:2004} on the persistence of the LASSO estimator. An interpretation of their results is that such an estimator is consistent.  However, there is a bias term of order $O(s \log p / n)$ inherent in the LASSO-based estimator, when the regularization parameter is optimally tuned.  When the bias is negligible, the LASSO based plug-in estimator is consistent.  The plug-in variance estimation based on the general folded-concave penalized least squares estimators such as SCAD are also discussed.
%We also use the relaxed LASSO proposed by \cite{Meinshausen:relaxedlasso:2007} to improve the convergence rate.
In some cases, this method is comparable with the RCV approach.

%\textbf{What can we do if the variance is known? Can the estimated variance give us some feedback about the model selection procedure?
%I mean, if the oracle tells us the true variance in advance, could we get more accurate result on variable selection? }

The paper is organized as following. Section \ref{sec2} gives some additional insights into the challenges of high dimensionality in variance estimation.
In Section \ref{sec3}, the RCV variance estimator is proposed and its sampling properties are established.  Section \ref{sec4} studies the variance estimation based penalized likelihood methods. Extensive simulation studies are conducted in Section \ref{sec5} to illustrate the advantage of the proposed methodology. Section \ref{sec:dis} is devoted to a discussion and the detailed proofs are provided in the Appendix.

\section{Insights into challenges of High Dimensionality in variance estimation} \label{sec2}

Consider the usual linear model
\begin{equation}  \label{a2}
Y_i = \bx_i^T \bbeta + \varepsilon_i, \quad \mbox{or} \quad
\mathbf{y}=\mathbf{X}\boldsymbol{\beta} + \boldsymbol{\varepsilon},
\end{equation}
where $\mathbf{y}=(Y_1,...,Y_n)^T$ is an $n$-vector of responses, $\mathbf{X}=(\mathbf{x}_1,...,\mathbf{x}_n)^{T}$ is an $n \times p$ matrix of independent
and identically distributed ($i.i.d.$) variables $\mathbf{x}_1$,..., $\mathbf{x}_n$, $\bbeta = (\beta_1,...,\beta_p)^T$ is a $p$-vector of parameters and $\boldsymbol\varepsilon = (\varepsilon_1,...,\varepsilon_n)^T$ is an $n$-vector of $i.i.d.$ random noises with mean 0 and variance $\sigma^2$.
We always assume the noise is independent of predictors.
For any index set $M\subset \{1,2,...,p\}$, $\bbeta_{M}$ denotes the sub-vector containing the components of the vector $\bbeta$ that
are indexed by $M$, $\mathbf{X}_{M}$ denotes the sub-matrix containing the columns of $\mathbf{X}$ that are indexed by $M$, and
$\mathbf{P}_{M}=\mathbf{X}_{M}(\mathbf{X}_{M}^T\mathbf{X}_{M})^{-1}\mathbf{X}_{M}^T$ is the projection operator onto the linear space generated by the column vectors of $\mathbf{X}_{M}$.

When $p>n$ or $p\gg n$, it is often assumed that the true model $M_0 = \{j: \beta_j \not = 0\}$ is sparse, i.e. the number of non-zero coefficients $s=|M_0|$ is small.
It is usually assumed that $s$ is fixed or diverging at a mild rate.
Under various sparsity assumptions and regularity conditions, the most popular variable selection tools such as LASSO,
SCAD, adaptive LASSO, SIS and Dantzig selector possess various good properties regarding model selection consistency.
%See also
%\cite{FanLv08,Fan:Lv:NP:2009,ISI:000245390700010,ISI:000258243000006,Zou:adap:2006,CandesTao07,bick:rito:simu:2008}.
Among these properties are the sure screening property, model consistency, sign consistency, weak oracle property and oracle property, from weak to strong.
Theoretically, under some regularity conditions, all aforementioned model selection tools can achieve model consistency. In other words, they
can exactly pick out the true sparse model with probability tending to one. However, in practice, these conditions are impossible to check and hard to meet.  Hence, it is often very difficult to extract the exact subset of
significant variables among a huge set of covariates. One of reasons is the spurious correlation, as we now illustrate.

Suppose that unknown to us the true data generating process in model (\ref{a2}) is
$$
     \bY = 2 \bX_1 + 0.3 \bX_2 + \beps
$$
where $\bX_j$ is the $n$-dimensional vector of the realizations of the covariate $X_j$.  Furthermore, let us assume that $\{X_j\}_{j=1}^p$ and $\varepsilon$ follow independently the standard normal distribution.  As illustrated in Figure 1(a), where $p$ is large, there are realizations of variables that have high correlations with $\beps$. Let us say $\widehat{\corr}(\bX_9, \beps) = 0.5$. Then, $X_{9}$ can even have a better chance to be selected than $X_2$.  Here and hereafter, we refer the spurious variables to those variables selected to predict the realized noise $\beps$ and their associated sample correlations are called spurious correlations.

Continued with the above example,  the naive-two stage estimator will work well when the model selection is consistent.  Since we may not get model consistency in practice and have no way to check even if we get it by chance, it is natural to ask whether the naive two-stage strategy works if only sure screening can be achieved in the first stage. In the aforementioned example, let us say a model selector chooses the set $\{X_1, X_2, X_9\}$, which contains all true variables.  However, in the naive two-stage fitting, $\bX_9$ is used to predict $\beps$, resulting in substantial underestimate of $\sigma^2 = \var(\varepsilon)$.  Upon both variables $X_1$ and $X_2$ are selected, all spurious variables are recruited to predict $\beps$.  The more spurious variables are selected, the better $\beps$ is predicted, the more serious underestimation of $\sigma^2$ by the naive two stage estimation.  %Let us formally define the meaning of sure screening.

%\textbf{Sure screening property.}
We say a model selection procedure satisfies sure screening property
if the selected model $\hat{M}$ with model size $\hat{s}$ includes the true model $M_0$ with probability tending to one. Explicitly,
$$P(\hat M \supset M_0)\to 1 \quad\text{as}\quad n \to \infty.$$

The sure screening property is a crucial criterion when evaluating a model selection procedure for high or ultrahigh dimensional problems.
Among all model consistent properties, the sure screening property is the weakest one and the easiest to achieve in practice.

Let us demonstrate the naive two-stage procedure in detail.
%In the high dimensional case, we can not apply the ordinary least squares method directly since the design matrix is always singular.
Assume that the selected model $\hat{M}$ in the first stage includes the true model $M_0$.
The ordinary least squares estimator $\hat{\sigma}^2_{\hat{M}}$ at the second stage,
using only the selected variables in $\hat M$, is
\begin{equation}
\hat{\sigma}^2_{\hat{M}} = \frac{\mathbf{y}^T(\mathbf{I}_n-\mathbf{P}_{\hat M})\mathbf{y}}{n-\hat{s}} = \frac{\beps^T(\mathbf{I}_n-\mathbf{P}_{\hat M})\beps}{n-\hat{s}}, \label{a3}
\end{equation}
where $\mathbf{I}_n$ is the $n\times n$ identity matrix.
How does this estimator perform?  %To understand its poor performance, suppose, for simplicity, that  $\bbeta=\bs{0}$ so that $\mathbf{y}= \beps$.
To facilitate the notation, denote the naive estimator by $\hat{\sigma}_n^2$.
Then, the estimator (\ref{a3}) can be written as
\begin{eqnarray*}
\hat{\sigma}^2_{n} = {1 \over n-\hat s}(1-\hat{\gamma}_n^2)\bs{\varepsilon}^T \bs{\varepsilon},
\end{eqnarray*}
where $\hat{\gamma}_n^2 = {\bs{\varepsilon}^T \bs{P}_{\hat M}\bs{\varepsilon} /
\bs{\varepsilon}^T\bs{\varepsilon}}.$ Let us analyze the asymptotic behavior of this naive two-stage estimator.

\begin{theorem} \label{thm1}  %{\bf Theorem 1.}
%Assume that $\bbeta=\bs{0}$.
Under the assumptions (A1)-(A2) together with (A3)-(A4) or (A5)-(A6) in the Appendix, we have
\begin{enumerate}
  \item If a procedure satisfies the sure screening property with $\hat{s}\leq b_n$ where $b_n = o(n)$ is given in Assumption (A2), then $\sigma_n^2/(1-\hat{\gamma}_n^2)$ converges to $\sigma^2$ in probability as $n\rightarrow \infty$. Furthermore,
\[
\sqrt{n}\Big\{\hat{\sigma}_n^2/(1-\hat{\gamma}_n^2) - \sigma^2\Big\} \stackrel{\mathcal{D}}{\to} N(0, E[\varepsilon_1^4]-\sigma^4),
\]
where `$\stackrel{\mathcal{D}}{\to}$' stands for `convergence in distribution'.
 % if $\sqrt{n}\hat{\gamma}_n^2 \rightarrow 0$, then
% \[
%\sqrt{n}(\hat{\sigma}_n^2 - \sigma^2) \stackrel{d}{\to} N(0, E[\varepsilon_1^4]-\sigma^4).
%\]
%\item If, $\log p /n \rightarrow 0$ and $\hat{s}= o_P(b_n)$, then $\hat{\gamma}_n = O_P(\sqrt{{\hat{s}\log p / n}})$.
\item If, in addition, $\log p /n = O(1)$, then $\hat{\gamma}_n = O_P(\sqrt{{\hat{s}\log p / n}})$.
\end{enumerate}
\end{theorem}

It is perhaps worthwhile to make a remark about Theorem~\ref{thm1}. $\hat{\gamma}_n^2$ plays an important role on the performance of $\hat{\sigma}_n^2$. It represents the fraction of bias in $\hat{\sigma}_n^2$.  The slower $\hat{\gamma}_n$ converges to zero, the worse $\hat{\sigma}_n^2$ performs. Moreover,
if $\hat{\gamma}_n^2$ converges to a positive constant with a non-negligible probability, it will lead to an inconstant estimator.  The estimator can not be root-n consistent if $\hat{s} \log p /\sqrt{n} \to \infty$.  This explains the poor performance of $\hat{\sigma}_n^2$, as demonstrated in Figures \ref{fig2} and \ref{fig3}.  While Theorem~\ref{thm1} gives an upper bound of $\gamma_n$, it is often sharp. For instance, if $\{X_j\}_{j=1}^p$ and $\varepsilon$ are $i.i.d.$ standard normal distribution and $\hat s = 1$, then $\hat{\gamma}_n$ is just the maximum absolute sample correlation between $\varepsilon$ and $\{X_j\}_{j=1}^p$. Denote the $j$th sample correlation by $\hat{\gamma}_{nj} = \widehat{\corr}_n(X_j, \varepsilon)$, $j = 1, \cdots,p$. Applying the transformation $T(r) = r/\sqrt{1 - r^2}$,
we get a sequence $\{\xi_{nj} = \sqrt{n-2}T(\hat{\gamma}_{nj})\}_{j=1}^p$ with $i.i.d.$ Student's $t$ distribution with $n-2$ degrees of freedom. Simple analysis on the extreme statistics of the sequences $\{\xi_{nj}\}$ and $\{\hat{\gamma}_{nj}\}$ shows that for any $c > 0$ such that
$\log (p/c) \leq n+2$, we have
\begin{eqnarray}  \label{a3a}
P\Big\{\hat{\gamma}_n >  \sqrt{\log (p/c)  / (2n)} \Big\} > 1 - \exp(-c),
\end{eqnarray}
which implies the sharpness of Theorem 1 in this specific case. Furthermore, when
$\log p=o(n^\frac12)$,
$$
    \hat{\gamma}_n = \sqrt{2 \log p / n} \{1 + o_p(1)\}
$$
with the limiting distribution is given by
\begin{equation}   \label{a3b}
P\Big\{\sqrt{2\log 2p}\big(\sqrt{n}\hat{\gamma}_n - d_{2p}\big) < x \Big\} \longrightarrow \exp\{-\exp(-x)\}.
\end{equation}
where $d_p = \sqrt{2\log p} - (\log\sqrt{ 4\pi\log p}) /\over \sqrt{2\log p}$.
See Appendix A.5 for details.

\section{Variance estimation based on refitted cross-validation}\label{sec3}
\subsection{Refitted cross-validation}

In this section, we introduce the refitted cross-validation method to remove the influence of spurious variables in the second stage. The method requires only that the model selection procedure in stage one has a sure screening property.
The idea is as follows. We assume the sample size $n$ is even for simplicity and split randomly the sample into two groups.
In the first stage, an ultrahigh dimensional variable selection method like SIS is applied to these two datasets separately,
which yields two small sets of selected variables.
In the second stage, the ordinary least squares (OLS) method is used to re-estimate the coefficient $\bbeta$ and variance $\sigma^2$. Different from the naive two-stage method, we apply again OLS to the \textbf{first subset} of the data with the variables selected by the \textbf{second subset} of the data and vice versa. Taking the average of these two estimators, we get our estimator of $\sigma^2$.  The refitting in the second stage is fundamental to reduce the influence of the spurious variables in the first stage of variable selection.

To implement the above idea of the refitted cross-validation, consider a dataset with sample size $n$, which is randomly split to two even datasets $(\mathbf{y}^{(1)}, \mathbf{X}^{(1)})$
and $( \mathbf{y}^{(2)}, \mathbf{X}^{(2)})$. First, a variable selection tool is performed on $(\mathbf{y}^{(1)}, \mathbf{X}^{(1)})$ and let $\hat M_1$
denote the set of selected variables. The variance  $\sigma^2$ is then estimated on the second dataset $(\mathbf{y}^{(2)}, \mathbf{X}^{(2)}_{\hat M_1})$, namely,
\[
\hat{\sigma}_1^2 = \frac{(\mathbf{y}^{(2)})^T(\mathbf{I}_{n/2}- \mathbf{P}^{(2)}_{\hat M_1})\mathbf{y}^{(2)}}{n/2-|\hat M_1|},
\]
where $\mathbf{P}^{(2)}_{\hat M_1} = \mathbf{X}^{(2)}_{\hat M_1}(\mathbf{X}^{(2)T}_{\hat M_1}\mathbf{X}^{(2)}_{\hat M_1})^{-1}\mathbf{X}^{(2)T}_{\hat M_1}$.
Similarly, we use the dataset two $( \mathbf{y}^{(2)}, \mathbf{X}^{(2)})$ to select the set of important variables $\hat M_2$ and the first dataset
$(\mathbf{y}^{(1)}, \mathbf{X}^{(1)}_{\hat M_2})$ for estimation of $\sigma^2$, resulting in
\[
\hat{\sigma}_2^2 = \frac{{(\mathbf{y}}^{(1)})^T(\mathbf{I}_{n/2}-\mathbf{P}^{(1)}_{\hat M_2})\mathbf{y}^{(1)}}{n/2-|\hat M_2|}.
\]
We define the final estimator as
\begin{equation} \label{a4}
\hat\sigma^2_{\sRCV} = (\hat\sigma_1^2 + \hat\sigma_2^2)/2.
\end{equation}
An alternative is the weighted average defined by
\begin{equation} \label{a5}
\hat\sigma^2_{\wRCV} = \frac{(\mathbf{y}^{(2)})^T(\mathbf{I}_{n/2}- \mathbf{P}^{(2)}_{\hat M_1})\mathbf{y}^{(2)}+
{(\mathbf{y}}^{(1)})^T(\mathbf{I}_{n/2}-\mathbf{P}^{(1)}_{\hat M_2})\mathbf{y}^{(1)}}{n-|\hat M_1|-|\hat M_2|}.
\end{equation}
When $|\hat M_1|=|\hat M_2|$, we have $\hat\sigma^2_{\sRCV}=\hat\sigma^2_{\wRCV}$.

%In the above procedure, although $\hat{M}_1$ includes some extra unimportant variables besides the important ones, these extra variables will play minor roles when we estimate $\sigma^2$ using the second dataset along with refitting. For example, if an irrelevant variable is selected in the $\hat{M}_1$ because its high correlation with the realized noise in the first dataset, then in the second dataset, its sample correlation with the realized noise is likely small.  Hence, it can hardly affect the estimation of $\sigma^2$.
In the above procedure, although $\hat{M}_1$ includes some extra unimportant variables besides the important ones, these extra variables will play minor roles when we estimate $\sigma^2$ using the second dataset along with refitting since they are just some random unrelated variables over the second dataset.  Furthermore, even when some important variables are missed in the first stage of model selection, they have a good chance being well approximated by the other variables selected in the first stage to reduce modeling biases. Thanks to the refitting in the second stage, the best linear approximation of those selected variables is used to reduce the biases.  Therefore, a larger selected model size gives us not only a better chance of sure screening, but also a way to reduce modeling biases in the second stage when some important variables are missing.  This explains why the RCV method is relatively insensitive to the selected model size, demonstrated in Figures 3 and 6 below. With a larger model being selected in the stage one, we may lose some degrees of freedom and hence get an estimator with slightly larger variance than the oracle one at finite sample. Nevertheless, the RCV estimator performs well in practice and asymptotically optimal when $\hat{s}=o(n)$.
The following theorem gives the property of the RCV estimator.  It requires only a sure screening property, studied by \cite{FanLv08} for normal multiple regression, \cite{FanSong2010} for generalized linear models, and \cite{ZhaoLi2010} for Cox regression model.

\begin{theorem}   \label{thm2}   %Theorem 2
Assume the regularity conditions (A1) and (A2) hold and $E[\varepsilon^4]<\infty$. If a procedure satisfies the sure screening property with $\hat{s}_1 \leq b_n$ and $\hat{s}_2 \leq b_n$, then
%($b_n/n \rightarrow 0$)
\begin{equation}  \label{a6}
\sqrt{n}(\hat\sigma^2_{\sRCV} - \sigma^2) \stackrel{\mathcal{D}}{\to} N(0, E[\varepsilon^4]-\sigma^4).
\end{equation}
\end{theorem}

Theorem~\ref{thm2} reveals that the RCV estimator of variance has an oracle property.  If the regression coefficient $\bbeta^*$ is known by oracle, then we can compute the realized noise $\varepsilon_i = Y_i - \bx_i^T \bbeta^*$ and get the oracle estimator
\begin{equation}  \label{a7}
    \hat{\sigma}_O^2 = n^{-1} \sum_{i=1}^n (Y_i - \bx_i^T \bbeta^*)^2.
\end{equation}
This oracle estimator has the same asymptotic variance as $\hat\sigma^2_{\sRCV}$.

There are two natural extensions of the aforementioned refitted cross-validation techniques.

\textbf{K-fold data splitting}:
The first natural extension is to use K-fold data splitting technique rather than two-fold spliting. We can divide the data into K groups, select the model with all
groups except one, which is used to estimate the variance with refitting. We may improve the sure screening probability with this K-fold method since there are now  more data in the first stage.  However, there are only $n/K$ data points on the second stage for refitting.  This means that the number of variables selected in the first stage should be much less than $n/K$.  This makes the ability of sure screening hard in the first stage.  For this reason, we work only on the two-fold refitted cross-validation.

\textbf{Repeated data splitting}:
There are many ways to randomly split the data.  Hence, many RCV variance estimators can be obtained.  We may take the average of the resulting estimators.
This reduces the influence of the randomness in the data splitting.  %Therefore, multiple data splitting is recommended for practical implementation.

{\bf Remark 1.} The RCV procedure provides an efficient method for variance estimation. Those technical conditions in Theorem 2 may not be weakest possible. They are imposed to facilitate the proofs. In particular, we assume that $ P\Big\{\phi_{\min}(b_n) \ge \lambda_0\Big\}=1$ for all $n$, which implies that the selected variables in stage one are not highly correlated. Other methods beyond least squares can be applied in the refitted stage when those assumptions are possibly violated in practice. For instance, if some selected variables in stage one are highly correlated or the selected model size is relatively large, ridge regression or penalization methods can be applied in the refitted stage. Moreover, if the density of the error $\varepsilon$ seems heavy-tailed, some classical robust methods can also be employed.

%Undoubtedly, several other classical methods beyond least squares can be applied in the refitted stage when those assumptions are possibly violated in practice. For instance, if some selected variables in stage one are highly correlated or the selected model size is relatively large, we prefer to adopt an alternative approach in the refitted stage, ridge regression method, to handle those difficulties, which is expected to perform better than least squares. Specifically, $\mathbf{P}_{M}$ is replaced by $\mathbf{\tilde{P}}_{M} = \mathbf{X}_{ M}(\mathbf{X}^{T}_{M}\mathbf{X}_{M} + \lambda_n \mathbf{I}_M)^{-1}\mathbf{X}^{T}_{M}$ in expressions (\ref{a4}) and (\ref{a5}) with $\lambda_n \to 0$. If the density of the error $\varepsilon$ seems heavy, some classical robust methods can also be used here.

{\bf Remark 2.} The paper focuses on variance estimation under the exact sparsity assumption and sure screening property.
%As described in Section 1,  these assumptions are common and often necessary for statistical inference in high or ultra-high dimensional setting.
It is possible to extend our results to nearly sparse cases. For example, the parameter $\bs{\beta}$ is not sparse but satisfies some decay condition such as
$ \sum_k | \bbeta_i | \leq C$
for some positive constant $C$. In this case, we do not have to worry too much whether a model selection procedure can recover small parameters. In this case, so long as a model selection method can pick up a majority of all variables with large coefficients in the first stage, we would expect that the RCV estimator performs well.

\subsection{Applications}
Many statistical problems require the knowledge of the residual variance, especially for high or ultra-high dimensional linear regression. Here we brief a couple of applications.

{\it (a) Constructing confidence intervals for coefficients.}  A natural application is to use estimated $\hat{\sigma}_{\sRCV}$ to construct confidence intervals for non-vanishing estimated coefficients. For example, it is well known that the SCAD estimator possesses an oracle property \citep{FanLi01, Fan:Lv:NP:2009}.  Let $\hat{\bbeta}_{\hat M}$ be the SCAD estimator, with corresponding design matrix $\bX_{\hat{M}}$.  Then, for each $j \in \hat{M}$, $1-\alpha$ confidence interval for $\beta_j$ is
\begin{equation}  \label{eqci}
   \hat{\beta}_j \pm z_{1-\alpha/2} c_j \hat{\sigma}_{\sRCV},
\end{equation}
in which $c_j$ is the diagonal element of the matrix $(\bX_{\hat{M}}^T \bX_{\hat{M}})^{-1}$ that corresponds to the $j^{th}$ variable.  Our simulation studies show that such a confidence interval is accurate and has a similar performance to the case where $\sigma$ is known.

The confidence intervals can also be constructed based on the raw materials in the refitted cross validation.  For example, for each element in $\hat{M} \equiv \hat{M}_1 \cap \hat{M}_2$, we can take the average of the refitted coefficients as the estimate of the regression coefficients in the set $\hat{M}$, and $(\bS_1 + \bS_2) \hat{\sigma}_{\sRCV}^2 /4$ as the corresponding estimated covariance matrix, where $\bS_1 = (\bX^{(1)}_{\hat{M}}{}^T \bX^{(1)}_{\hat{M}})^{-1}$ is computed based on the first half of the data at the refitting stage and $\bS_2 = (\bX^{(2)}_{\hat{M}}{}^T \bX^{(2)}_{\hat{M}})^{-1}$ is computed based on the second half of the data.  In addition, some `cleaning' techniques through $p$-values can be also applied here. In particular, \citet{Wasserman:Roeder:HDVS:2009} and \cite{MMB09} studied these techniques to reduce the number of falsely selected variables substantially.

{\it (b) Genomewide association studies}.  Let $X_j$ be the coding of the $j$th Single Nucleotide Polymorphism (SNP) and $Y$ be the observed phenotype (e.g. height or blood pressure) or the expression of a gene of interest.  In such a quantitative trait loci (QTL) or eQTL study, one frequently fits the marginal linear regression
\begin{equation}  \label{qtl}
    E(Y|X_j) = \alpha_j + \beta_j X_j
\end{equation}
based on a sample of size $n$ individuals, resulting in the marginal least-squares estimate $\hat{\beta}_j$.  The interest is to test simultaneously the hypotheses $H_{0, j}: \beta_j = 0$ $(j=1, \cdots, p)$. If the conditional distribution of $Y$ given $X_1, \cdots, X_p$ is $N(\mu(X_1, \cdots, X_p), \sigma^2)$, then it can easily be shown \citep{HanGuFan10} that $(\hat \beta_1, \cdots, \hat \beta_p)^T \sim N((\beta_1, \cdots, \beta_p)^T, \sigma^2 \bS/n)$, where the $(i,j)$ element of $\bS$ is the sample covariance matrix of $X_i$ and $X_j$ divided by their sample variances.  With $\sigma^2$ estimated by the RCV, the P-value for testing individual hypothesis $H_{0, j}$ can be computed.  In addition, the dependence of the least-squares estimates is now known and hence the false discovery proportion or rate can be estimated and controlled \citep{HanGuFan10}.

{\it (c) Model selection.} Popular penalized approaches for variable selection such as LASSO, SCAD, adaptive LASSO and elastic-net often involve the choice of tuning or regularization parameter. A proper tuning parameter can improve the efficiency and accuracy for variable selection. Several criteria, such as Mallaw's $\mathcal{C}_p$, AIC and BIC, are constructed to choose tuning parameters. All these criteria rely heavily on a common parameter, the error variance. As an illustration, consider estimating the tuning parameter of LASSO (See also \cite{ZHT07}). Let $\lambda$ be the tuning parameter with the fitted value $\bs{\hat{\mu}}_{\lambda} =\mathbf{X}\bs{\hat{\beta}}_{\lambda}$. Then AIC and BIC for the LASSO are written as
 $$
 \mbox{AIC}(\bs{\hat{\mu}}_{\lambda},\sigma^2) = {||\mathbf{y} - \bs{\hat{\mu}}_{\lambda}||^2 \over n \sigma^2} + {2 \over n} \widehat{df}(\bs{\hat{\mu}}_{\lambda})
 $$
 and
 $$
 \mbox{BIC}(\bs{\hat{\mu}}_{\lambda},\sigma^2) = {||\mathbf{y} - \bs{\hat{\mu}}_{\lambda}||^2 \over n \sigma^2} + {\log (n) \over n} \widehat{df}(\bs{\hat{\mu}}_{\lambda}).
 $$
 It is easily seen that the variance $\sigma^2$ has an important impact on both AIC and BIC. %With the good estimator $\hat{\sigma}^2_{\sRCV}$,  the performance of AIC and BIC is expected to improve, especially for high dimensional settings.

\section{Folded-concave Penalized least squares}\label{sec4}
In this section, we discuss some related methods on variance estimation and their corresponding asymptotic properties.  The oracle estimator of $\sigma^2$ is
$$
  \hat{R}(\bbeta^*)=n^{-1} \sum_{i=1}^{n} \left(Y_i-\mathbf{x}_i^T \bbeta^*\right)^2
$$
A natural candidate to estimate the variance is $\hat{R}(\hat{\bbeta})$, where $\hat{\bbeta}$ is the LASSO or SCAD estimator of $\bbeta^*$.
\citet{Greenshtein:Ritov:persistence:2004} showed the persistent property for the LASSO estimator $\hat{\bbeta}_{L}$.
Their result, interpreted in the linear regression setting, implies
$R(\hat{\bbeta}_{L})\rightarrow R(\bbeta^*)=\sigma^2$ in probability, where
$R(\bbeta) = E(Y - \bX \bbeta)^2$. In fact, it is easy to see that their result implies
$$
   \hat{R}(\hat{\bbeta}_{L})\rightarrow \sigma^2 = R(\bbeta^*).
$$
In other words, $\hat{R}(\hat{\bbeta}_{L})$ is a consistent estimator for the variance.

Recall the LASSO estimator is defined as
\begin{equation}
\label{a8}
\hat{\bbeta}_{L}= \argmin\limits_{ \bbeta }
\frac1n \sum_{i=1}^{n} \left(Y_i-\mathbf{x}_i^T \bbeta \right)^2 + \lambda_n\|\bbeta\|_{1}.
\end{equation}
To make $\hat{R}(\hat{\bbeta}_{L})$ consistent, \citet{Greenshtein:Ritov:persistence:2004}
suggested $\lambda_n=o\{(n/\log p)^\frac12\}$ asymptotically. \citet{Wasserman:Roeder:HDVS:2009} showed the consistency still holds when $\lambda_n$ is chosen by cross-validation. Therefore, we define the LASSO variance estimator $\hat{\sigma}^2_{L}$ by
\begin{equation}\label{a9}
\hat{\sigma}^2_{L}={1 \over n - \hat{s}_{L}} \sum_{i=1}^{n} \left(Y_i-\mathbf{x}_i^T\hat{\bbeta}_{L}\right)^2,
\end{equation}
where $\hat{s}_{L} = \#\{j: (\hat{\beta}_{L})_{j}\neq 0\}$.

We shall see that $\hat{\sigma}^2_{L}$ usually underestimates the variance due to spurious correlation, as the LASSO shares a similar spirit of the stepwise addition (see the LARS algorithm by \cite{EHJT04}). Thus, we also consider the leave-one-out LASSO variance estimator
\begin{equation}
\label{a10}
\hat{\sigma}^2_{\sLL}=\frac{1}{n} \sum_{i=1}^{n} \left(Y_i-\mathbf{x}_i^T\hat{\bbeta}^{(-i)}_{L}\right)^2
\end{equation}
where $\hat{\bbeta}^{(-i)}_{L}$ is the LASSO estimator using all samples except the $i$th one.
In practice, $K$-fold ($K$ equals $5$ or $10$) cross-validated LASSO estimator is often used and shares the same spirit as (\ref{a10}). We divide the dataset into $K$ parts, say $\mathcal{D}_1,...,\mathcal{D}_K$ and define
\begin{equation}\label{a11}
\hat{\sigma}^2_{\sCVL}=\min\limits_{\lambda}\frac{1}{n} \sum_{k=1}^{K}\sum_{i\in\mathcal{D}_k} \left(Y_i-\mathbf{x}_i^T\hat{\bbeta}^{(-k)}_{\lambda}\right)^2
\end{equation}
where $\hat{\bbeta}^{(-k)}_{\lambda}$ is the LASSO estimator using all data except ones in $\mathcal{D}_{k}$ with tuning parameter $\lambda$.  This estimator differs from the plug-in method (\ref{a9}) in that multiple estimates from training samples are used to compute residuals from the testing samples.  We will see that the estimator $\hat{\sigma}^2_{\sCVL}$ is typically closer to $R(\hat{\bbeta}_{L})$ than $\hat{R}(\hat{\bbeta}_{L})$, but it usually somewhat overestimates the true variance from our simulation experience. The following theorem shows the convergence rate for the LASSO estimator.

\begin{theorem} \label{thm3}   %Theorem 3
Suppose the assumptions (A1) - (A4) and (A7) hold. If the true model size $s = o(n^{\alpha_0})$ for some $\alpha_0 < 1$, then, we have
$$
\hat{\sigma}_L^2 - \sigma^2 = O_P(\max(n^{-1/2}, s \log p/n)).
$$
If $s \log p /\sqrt{n} \to 0$, we have
$$
     \sqrt{n} (\hat{\sigma}_L^2 - \sigma^2) \to N(0, E [\varepsilon^4] - \sigma^4).
$$
\end{theorem}

\vspace{2ex}
The factor $s \log p / n$ reflects the bias of the penalized $L_1$-estimator.  It can be non-negligible.  When it is negligible, the plug-in LASSO estimator possesses also the oracle property. In general, it is difficult to study the asymptotic distribution of the LASSO estimator when the bias is not negligible.
In particular, we can not obtain the standard error for the estimator. Even for finite $p$, \cite{KnightFu00} investigated the asymptotic distribution of LASSO-type estimators but it is too complicated to be applied for inference. To tackle this difficulty, \cite{ParkCasella08} and \cite{MGMC2010} used the hierarchical Bayesian formulation to produce a valid standard error for LASSO estimator, and \cite{ChatterjeeLahiri10} proposed a modified bootstrap method to approximate the distribution of LASSO estimator. But it is unclear yet whether or not their methods can be applied to high or ultra-high dimensional setting.

Recently, \citet{Fan:Lv:NP:2009} studied the oracle properties of non-concave penalized likelihood method in the ultrahigh dimensional setting. Inspired by their results, the variance $\sigma^2$  can be consistently and efficiently estimated.
The SCAD penalty $\rho_{\lambda}(t)$ \citep{FanLi01} is the function whose derivative is given by
$$
\rho'_{\lambda}(t) = \lambda\big\{\mbox{I}(t \le \lambda) + {(a\lambda - t)_+ \over (a-1) \lambda }\mbox{I}(t > \lambda)\big\}, t \ge 0, a >2,
$$
where $a = 3.7$ is often used.
Denote by
\begin{eqnarray}
\label{a12}
&&\bs{Q}_{n,\lambda_n}(\bbeta) = \| \mathbf{y}- \bX \bbeta \|^2 + 2n\sum_{j=1}^p\rho_{\lambda_n}(|\beta_j|),
\end{eqnarray}
and let $ \hat{\bs \beta}_{\sSCAD}$ be a local minimizer of $\bs{Q}_{n,\lambda_n}(\bbeta)$ with respect to $\bbeta$.
Thus, the variance $\sigma^2$  can be estimated by
$$
\hat{\sigma}_{\sSCAD}^2 = {1 \over {n-\hat{s}}} \sum_{i=1}^n(Y_i-\mathbf{x}_i^T\hat{\bs \beta}_{\sSCAD})^2,
$$
where $\hat{s} = \#\{j: (\hat{\beta}_{\sSCAD})_{j}\neq 0\}$.

The following theorem shows the oracle property and convergence rate for the SCAD estimator.

\begin{theorem}   \label{thm4}  %Theorem 4.
Assume $\log p = O(n^{\alpha_0})$ and the true model size $s = O(n^{\alpha_0})$, where $\alpha_0 \in [0,1)$.  Suppose that the assumptions
(A1), (A3)-(A4) (or (A5)-(A6)) and (A8)-(A9) in the Appendix are satisfied. Then,
\begin{enumerate}
\item (Model Consistency) There exists a strictly local minimizer $\hat{\bbeta}_n = (\hat{\beta}_1,\cdots,\hat{\beta}_p)^T$ of $\bs{Q}_{n,\lambda_n}(\bbeta)$ such that
$$
\{j: \hat{\beta}_j \neq 0\} = M_0
$$
with probability tending to one;
\item (Asymptotic Normality) With this estimator $\hat{\bbeta}_n$, we have
$$
\sqrt{n} \left ( \hat{\sigma}^2_{\sSCAD} - \sigma^2 \right ) \stackrel{\mathcal{D}}{\longrightarrow} N(0,E[\varepsilon^4] - \sigma^4).
$$
\end{enumerate}
\end{theorem}

Theorem~\ref{thm4} reveals that, if $\lambda_n$ is chosen reasonably, $\hat{\sigma}^2_{\sSCAD}$ works as well as the RCV estimator $\hat{\sigma}^2_{\sRCV}$ and better than $\hat{\sigma}^2_{L}$. However, it is hard to achieve this oracle property sometimes.

%%%%%%%%%%%%%%%%%%%%%%%%%%%%%%%%%%%%%%%%%%%%%%%%%%%%%%%%%%%%%%%%%%%%%%%%
\section{Numerical Results}\label{sec5}
\subsection{Simulation Study}\label{sec:sim1}
In this section, we illustrate and compare the finite sample performance of the methods described in the last three sections.
We applied these methods to three examples: the null model and two sparse models.
The null model (Example 1) is given by
\begin{equation}   \label{a13}
    Y= \bx^T \mbox{\bf 0} + \varepsilon, \quad \varepsilon \sim N(0, 1)
\end{equation}
where $X_1$, $X_2$, $\cdots$, $X_p$ are $i.i.d.$ random variables, following the standard Gaussian distribution.
This is the sparsest possible model.  The second sparse model (Example 2) is given by
\begin{equation}  \label{a14}
    Y=b(X_1+X_2+X_3)+\varepsilon, \quad \varepsilon \sim N(0, 1)
\end{equation}
with different $b$ representing different levels of signal-to-noise ratio (SNR).
The covariates associated with model (\ref{a14}) are jointly normal with equal correlation $\rho$, and marginally $N(0,1)$.
%$$
%    \corr(X_i,X_j)=\rho \quad \forall i\neq j.
    %\mbox{ $\forall i<j\leq q$ and }
%    \corr(X_i,X_j)=0 \mbox{ otherwise }.
%$$
%We consider $i.i.d.$ case and correlation case $\rho=0.5$ and $q=50$.

The third sparse model (Example 3) is more challenging, with 10 nontrivial coefficients, $\{\beta_j \mid j=1,2,3,5,7,11,13,17,19,23\}$. The covariates are jointly normal with $\cov(X_i,X_j)=0.5^{|i-j|}$. %The nonzero coefficients are generated from double exponential distribution $p(x)=\exp\{-2|x|\}$ and then rescaled to fit different SNR levels.
The nonzero coefficients vector is $$b\cdot(1.01,-0.06,0.72,1.55,2.32,-0.36,3.75,-2.04,-0.13,0.61)$$ where b varies to fit different SNR levels.
The random error follows the standard normal distribution.

In each of these settings, we test the following four methods to estimate the variance.

\textbf{Method 1}: Oracle estimator (\ref{a7}), which is not a feasible estimator whose performance provides a benchmark.

\textbf{Method 2}: Naive two-stage method, denoted by \textbf{N-SIS}, if SIS is employed in the model selection step.

\textbf{Method 3}: Refitted cross-validation variance estimator (\ref{a4}), denoted by \textbf{RCV}.

\textbf{Method 4}: One step method via penalized least squares estimators. We introduced this method in Section \ref{sec4} and recommended two formulas to estimate the variance,
direct plug-in method (\textbf{P}) like formula (\ref{a9}) and cross-validation method (\textbf{CV}) like formula (\ref{a11}).

In methods 2--4, we employed (I)SIS, SCAD, LASSO as our model selection tools. For SCAD and LASSO, the tuning parameters were chosen by 5-fold or 10-fold cross-validation. For (I)SIS, the predetermined model size is always taken to be 5 in the null model and $n/4$ in the sparse model, unless specified explicitly.  The principled method of \cite{ZhaoLi2010} can be employed to automatically choose the model size.

%The bias (\textbf{BIAS}) and the standard error (\textbf{SE}) of each estimator are reported. The average model size (\textbf{AMS}) and sure screening probability (\textbf{SSP}) of selected models are also helpful for us to understand the behaviors of our estimators.  They are reported too.

\begin{figure}[htbp]
\includegraphics[width=72mm,height=162mm,angle=270]{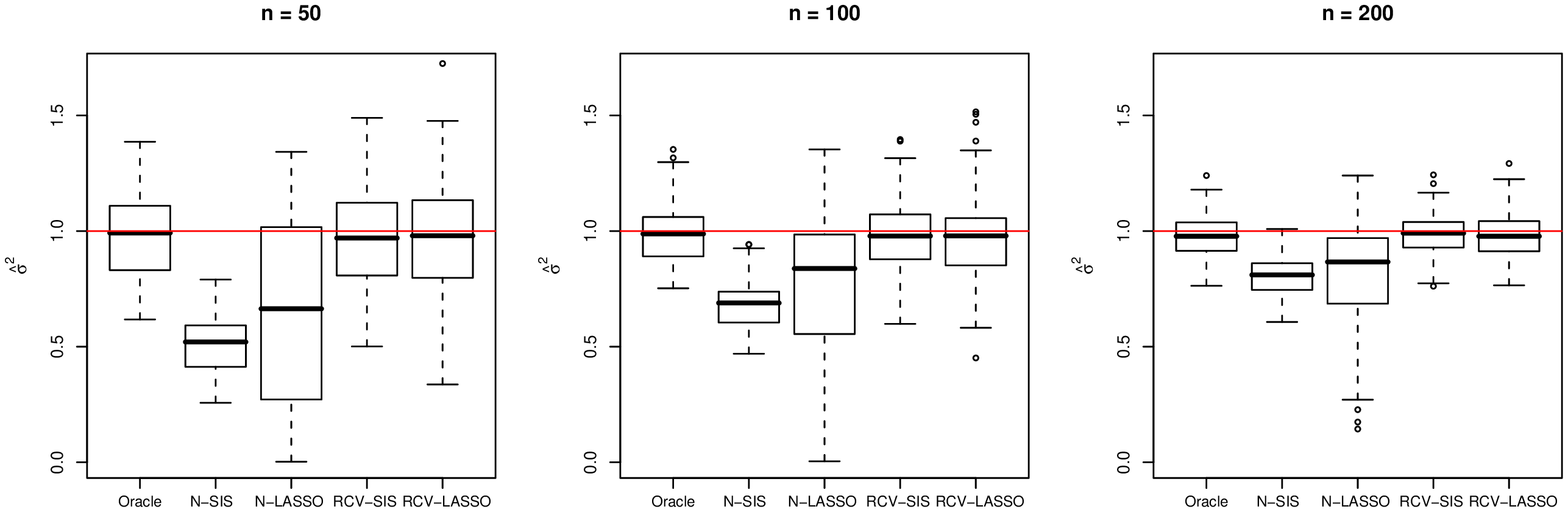}
\caption{Boxplots of $\hat{\sigma}_n^2$ when data are generated from the null model (\ref{a13}) with $p=1000$ and $n = 50, 100$ and 200.  The number of simulation is 100.  The horizontal line marks the true variance 1.}\label{fig4}
\end{figure}

{\bf Example 1.}
Assume the response $Y$ is independent of all predictors $X_i$'s, which follow $i.i.d.$ standard Gaussian distribution. We consider the cases when numbers of covariates vary from 10, 100 to 1000 and the sample sizes equal 50, 100 and 200. The simulation results are based on 100 replications and summarized in Table 1. In Figure \ref{fig4}, three boxplots are listed to compare the performance of different methods for the case $n=50, 100, 200$ and $p=1000$. From the simulation results, we can see the improved two-stage estimators (RCV-SIS and RCV-LASSO) are comparable with the oracle estimator and much better than the naive ones, especially in the case when $p \gg n$. This coincides with our theoretical result.
RCV improves dramatically the naive (natural) method, no matter SIS or LASSO is used.

%%%%%%%%%%%%%%%%%%%%%%%%%%%%%%%%%%%%%%%%%%%%%%%%%%%%%%%%%%%%%%%%%%%%%
%
%                     table of Example 1
%
%%%%%%%%%%%%%%%%%%%%%%%%%%%%%%%%%%%%%%%%%%%%%%%%%%%%%%%%%%%%%%%%%%%%%
\begin{table}[htbp]\scriptsize
\caption{Simulation Results for Example 1: The bias (BIAS), Standard Error (SE) and Average Model Size (AMS) for oracle, naive and RCV two-stage procedures are reported below.}
\centering
 \begin{tabular}{|l|rcr|rcr|rcr|}
 \hline \hline
 \multicolumn{10}{c}{$p=10$}   \\ \hline
 \hline            &       & $n=50$ &      &   & $n=100$ &         &    &$n = 200$ &  \\
 \hline  Method   &  BIAS & SE  & AMS    &  BIAS & SE  & AMS    & BIAS & SE    & AMS\\
  \hline \hline
  Oracle          & 0.006 & 0.220 & 0    & -0.023 & 0.144 & 0   &-0.015 & 0.109 & 0 \\
  \hline
  N-SIS      & -0.072 & 0.209 & 5    & -0.064 & 0.142 & 5    &-0.030 & 0.109 & 5 \\
  RCV-SIS & 0.017 & 0.234 & 5     & -0.029 & 0.150 & 5    &-0.013 & 0.114 & 5 \\
  \hline
  N-LASSO    & -0.052 & 0.211 & 1.08 & -0.051 & 0.148 & 1.01 &-0.028 & 0.108 & 0.94 \\
  RCV-LASSO& -0.003 & 0.219 & 1.41  & -0.026 & 0.149 & 1.24 &-0.015 & 0.110 & 1.02 \\
   \hline \hline
 \multicolumn{10}{c}{$p=100$}   \\
 \hline
 \hline           &       & $n=50$ &      &   & $n=100$ &         &    &$n = 200$ &  \\
 \hline  Method   &  BIAS & SE  & AMS    &  BIAS & SE  & AMS    & BIAS & SE    & AMS\\
  \hline \hline
  Oracle          & -0.011 & 0.205 & 0    & 0.023 & 0.154 & 0    &-0.010 & 0.154 & 0 \\
  \hline
  N-SIS      & -0.325 & 0.151 & 5    & -0.164 & 0.135 & 5    &-0.112 & 0.135 & 5 \\
  RCV-SIS & -0.004 & 0.216 & 5     & 0.018 & 0.165 & 5     &-0.009 & 0.165 & 5 \\
  \hline
  N-LASSO    & -0.272 & 0.319 & 5.90 & -0.153 & 0.279 & 13.56 &-0.073 & 0.279 & 3.16 \\
  RCV-LASSO &  0.032 & 0.359 & 4.67 & 0.022 & 0.171 & 5.89   &-0.010 & 0.171 & 12.41 \\
% \end{tabular}

\hline \hline
 \multicolumn{10}{c}{$p=1000$}   \\ \hline
 \hline           &       & $n=50$ &      &   & $n=100$ &         &    &$n = 200$ &  \\
 \hline  Method   &  BIAS & SE  & AMS    &  BIAS & SE  & AMS    & BIAS & SE    & AMS\\
  \hline \hline
  Oracle          & -0.011 & 0.176 & 0    & -0.015 & 0.130 & 0   &-0.015 & 0.095 & 0 \\
  \hline
  N-SIS      & -0.488 & 0.118 & 5    & -0.314 & 0.098 & 5    &-0.192 & 0.079 & 5 \\
  RCV-SIS & -0.017 & 0.211 & 5     & -0.018 & 0.144 & 5    &-0.012 & 0.098 & 5 \\
  \hline
  N-LASSO    & -0.351 & 0.399 & 7.47 & -0.256 & 0.330 & 9.37 &-0.196 & 0.251 & 9.90 \\
  RCV-LASSO & -0.029 & 0.266 & 5.03 & -0.022 & 0.186 & 8.27 &-0.014 & 0.103 & 8.79 \\
   \hline \hline
 \end{tabular}
\end{table}

%%%%%%%%%%%%%%%%%%%%%%%%%%%%%%%%%%%%%%%%%%%%%%%%%%%%%%%%%%%%%%%%%%%%
%
%                     Boxplot of Example 1
%
%%%%%%%%%%%%%%%%%%%%%%%%%%%%%%%%%%%%%%%%%%%%%%%%%%%%%%%%%%%%%%%%%%%%%
%\begin{figure}
%\includegraphics[width=162mm,height=80mm]{figures/ex1p1000.eps}
%\caption{Boxplots of $\hat{\sigma}_n^2$ when $p=1000$ based on $100$ simulations.}
%\end{figure}

\begin{figure}[htbp]
\includegraphics[width=80mm,height=162mm,angle=270]{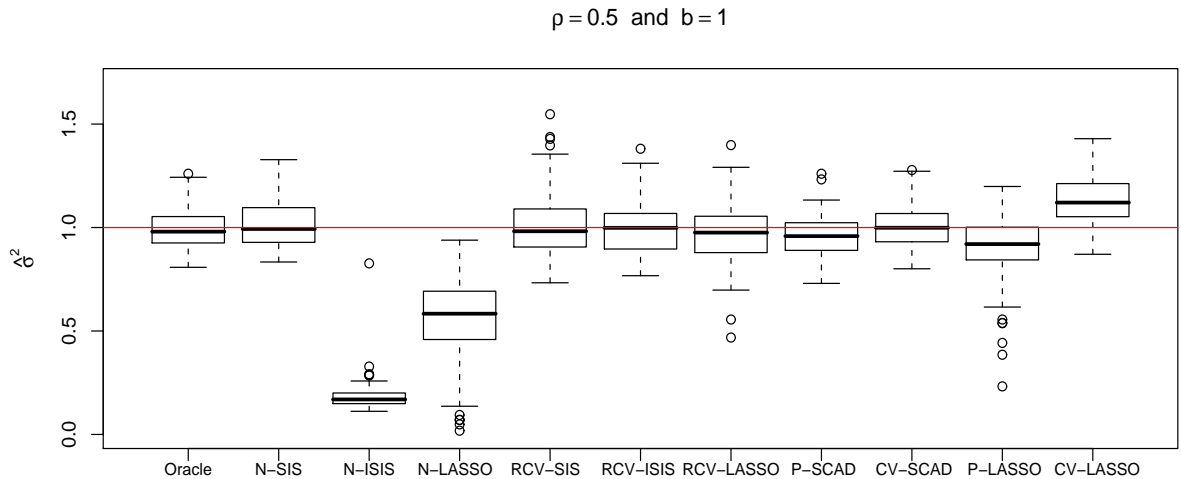}
\caption{Comparison of various methods for variance estimation in model (\ref{a14}) with $n=200$ and $p=2000$.  Presented are boxplots of $\hat{\sigma}_n^2$ based on $100$ replications.}\label{fig5}
\end{figure}

%%%%%%%%%%%%%%%%%%%%%%%%%%%%%%%%%%%%%%%%%%%%%%%%%%%%%%%%%%%%%%%%%%%%%%%%%%%%%%
%
%                                Example 2
%
%%%%%%%%%%%%%%%%%%%%%%%%%%%%%%%%%%%%%%%%%%%%%%%%%%%%%%%%%%%%%%%%%%%%%%%%%%%%%%%
{\bf Example 2.} % We now consider the model (\ref{a14}) under the two different configurations: $i.i.d.$ case $\rho=q=0$ and correlated case $\rho=0.5$, $q=50$. $(n,p)$ is set to be $(200,2000)$.
We now consider the the model (\ref{a14}) with $(n,p)=(200,2000)$, $\rho=0$ and $0.5$.
Moreover, we consider three values of coefficients $b=2$, $b=1$ and $b=1/\sqrt{3}$, corresponding to different levels of SNR $\sqrt{12}$, $\sqrt{3}$ and $1$ for each case when $\rho = 0$.  The results depicted in Table 2 are based on 100 replications (The results for $b=1$ are presented in Figure \ref{fig5} and are omitted from the table). The boxplots of all estimators for the case $\rho=0.5$ and $b=1$ are shown in Figure \ref{fig5}.  They indicate that the RCV methods behave as well as oracle, and much better than naive two-stage methods. Furthermore, the performance of the naive two-stage method depends highly on the model selection technique. The one-step methods perform also well, especially P-SCAD and CV-SCAD. P-LASSO and CV-LASSO behave slightly worse than SCAD methods. These simulation results lend further support to our theoretical conclusions in earlier sections.

\begin{figure}[htbp]
\includegraphics[width=80mm,height=162mm,angle=270]{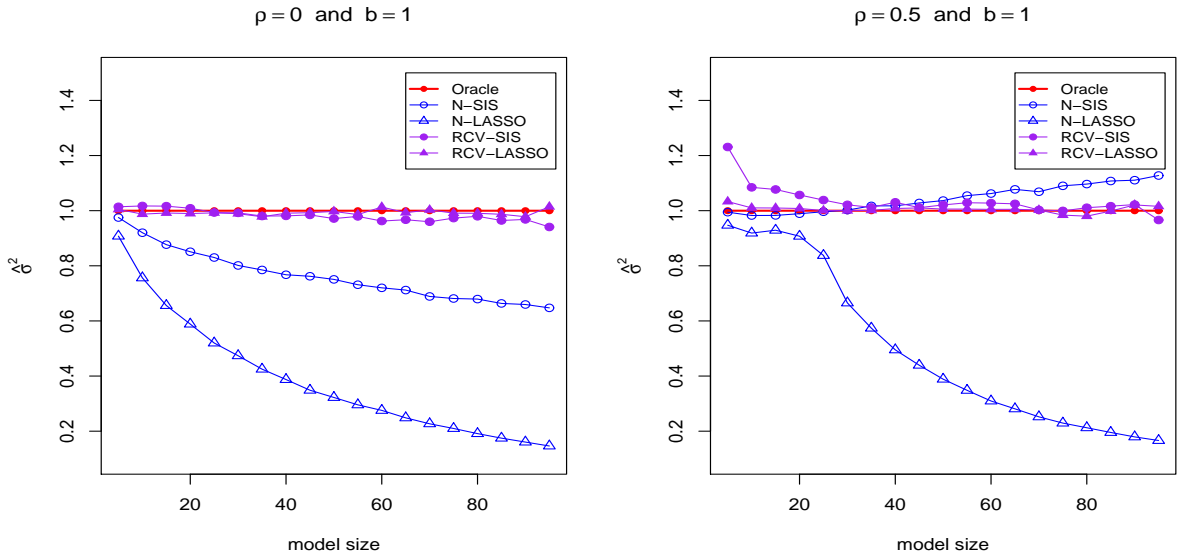}
\caption{The sensitivity of model size $\hat{s}$ on variance estimation. Presented are the medians of naive and RCV two-stage estimators when $n=200$ and $p=2000$ among $100$ replications.}\label{fig5.1}
\end{figure}

\begin{figure}[htbp]
\includegraphics[width=80mm,height=162mm,angle=270]{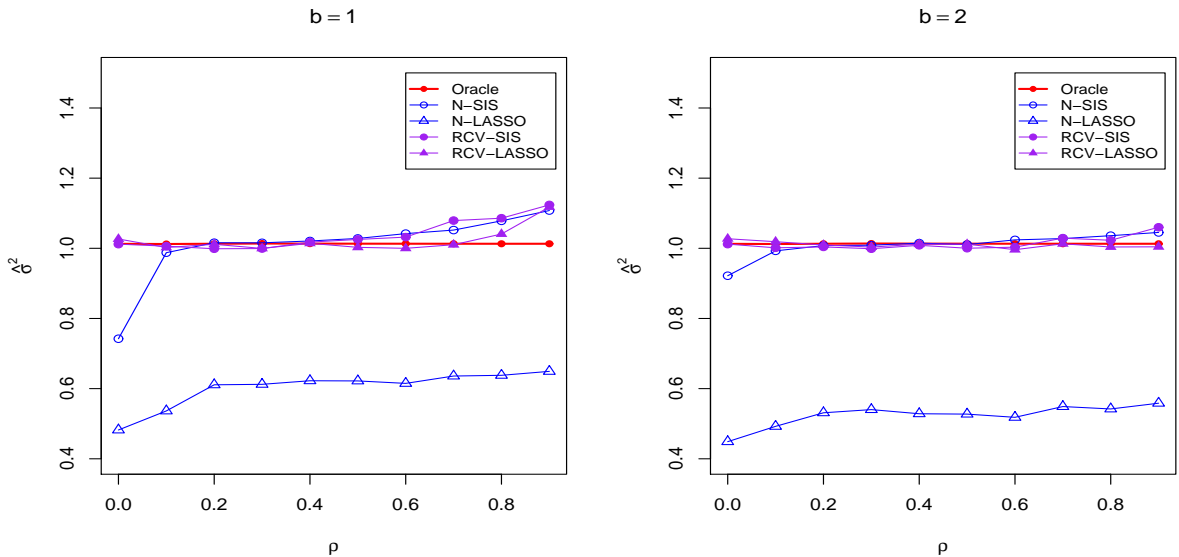}
\caption{The impact of covariance structure on variance estimation.  Presented are the medians of naive and RCV two-stage estimators when $n=200$ and $p=2000$ among $100$ replications for various $\rho$.}\label{fig5.2}
\end{figure}

To test the sensitivity of the RCV procedure to the model size $\hat{s}$ and covariance structure among predictors, additional simulations have been conducted and their results are summarized in Figure \ref{fig5.1} and \ref{fig5.2}. From Figure \ref{fig5.1}, it is clear that RCV method is insensitive to model size $\hat{s}$, as explained before Theorem 2. Figure \ref{fig5.2} shows the RCV methods are also robust with respect to the covariance structure. In contrast, N-LASSO always underestimates the variance.

\begin{table}[htbp]\scriptsize
\caption{Simulation results for Example 2 with $n=200$, $p=2000$: The bias (BIAS), Standard Error (SE), Average Model Size (AMS) and Sure Screening Probability (SSP) for each procedure are reported.} \label{tableEx2-3}
\centering
\begin{tabular}{|l|rcrc|rcrc|}
 \hline \hline
 \multicolumn{9}{c}{$b=2 $}   \\ \hline
 \hline   & & $\rho=0$ & & & & $\rho=0.5$ & & \\
 \hline  Method   &  BIAS & SE  & AMS  &  SSP  &  BIAS & SE  & AMS  &  SSP \\
  \hline \hline
  Oracle & -0.014 & 0.089 & 3.000 & 1.000 & -0.014 & 0.090 & 3.000 & 1.000 \\
  \hline
  N-SIS & -0.111 & 0.096 & 50.000 & 1.000 & -0.011 & 0.102 & 50.000 & 1.000 \\
  N-ISIS & -0.791 & 0.073 & 49.130 & 1.000 & -0.821 & 0.036 & 46.870 & 1.000 \\
  N-LASSO & -0.581 & 0.163 & 41.460 & 1.000 & -0.526 & 0.172 & 43.310 & 1.000 \\
  \hline
  RCV-SIS & -0.030 & 0.132 & 50.000 & 1.000 & 0.025 & 0.279 & 50.000 & 0.960 \\
  RCV-ISIS & -0.017 & 0.113 & 25.770 & 1.000 & -0.020 & 0.106 & 22.185 & 1.000 \\
  RCV-LASSO & -0.004 & 0.130 & 34.230 & 1.000 & -0.026 & 0.147 & 34.990 & 1.000 \\
  \hline
  P-SCAD  & -0.048 & 0.109 & 7.810 & 1.000 & -0.036 & 0.097 & 6.080 & 1.000 \\
  CV-SCAD & 0.000 & 0.095 & 7.810 & 1.000 & 0.001 & 0.096 & 6.080 & 1.000 \\
  P-LASSO & -0.102 & 0.195 & 41.460 & 1.000 & -0.113 & 0.164 & 43.310 & 1.000 \\
  CV-LASSO & 0.141 & 0.111 & 41.460 & 1.000 & 0.127 & 0.116 & 43.310 & 1.000 \\
%  \hline \hline
% \multicolumn{9}{c}{$b=1 $}   \\ \hline
% \hline   & & $\rho=0$ & & & & $\rho=0.5$ & & \\
% \hline  Method   &  BIAS & SE  & AMS  &  SSP  &  BIAS & SE  & AMS  &  SSP \\
%  \hline \hline
%  Oracle & -0.014 & 0.089 & 3.000 & 1.000 & -0.014 & 0.090 & 3.000 & 1.000 \\
%  \hline
%  N-SIS & -0.260 & 0.087 & 50.000 & 1.000 & 0.010 & 0.105 & 50.000 & 1.000 \\
%  N-ISIS & -0.782 & 0.096 & 48.770 & 1.000 & -0.817 & 0.077 & 46.400 & 1.000 \\
%  N-LASSO & -0.551 & 0.178 & 38.450 & 1.000 & -0.445 & 0.202 & 39.290 & 1.000 \\
%  \hline
%  RCV-SIS & -0.008 & 0.147 & 50.000 & 0.980 & 0.017 & 0.164 & 50.000 & 0.880 \\
%  RCV-ISIS & -0.017 & 0.103 & 27.145 & 1.000 & -0.002 & 0.122 & 22.225 & 0.970 \\
%  RCV-LASSO & -0.006 & 0.173 & 32.760 & 1.000 & -0.029 & 0.147 & 33.470 & 0.990 \\
%  \hline
%  P-SCAD & -0.054 & 0.106 & 9.270 & 1.000 & -0.036 & 0.096 & 6.110 & 1.000 %\\
%  CV-SCAD & 0.006 & 0.098 & 9.270 & 1.000 & 0.003 & 0.096 & 6.110 & 1.000 %\\
%  P-LASSO & -0.089 & 0.199 & 38.450 & 1.000 & -0.097 & 0.171 & 39.290 & 1.000 \\
%  CV-LASSO & 0.141 & 0.111 & 38.450 & 1.000 & 0.126 & 0.116 & 39.290 & 1.000 \\
 \hline \hline
 \multicolumn{9}{c}{$b=1/\sqrt{3}$}   \\ \hline
 \hline   & & $\rho=0$ & & & & $\rho=0.5$ & & \\
 \hline  Method   &  BIAS & SE  & AMS  &  SSP  &  BIAS & SE  & AMS  &  SSP \\
  \hline \hline
 Oracle & -0.014 & 0.090 & 3.000 & 1.000 & -0.014 & 0.090 & 3.000 & 1.000 \\
  \hline
  N-SIS & 0.010 & 0.105 & 50.000 & 1.000 & 0.046 & 0.107 & 50.000 & 0.980 \\
  N-ISIS & -0.817 & 0.077 & 46.400 & 1.000 & -0.809 & 0.099 & 46.250 & 1.000 \\
  N-LASSO & -0.445 & 0.202 & 39.290 & 1.000 & -0.381 & 0.239 & 37.140 & 1.000 \\
  \hline
  RCV-SIS & 0.017 & 0.164 & 50.000 & 0.880 & 0.057 & 0.158 & 50.000 & 0.430 \\
  RCV-ISIS & -0.002 & 0.122 & 22.225 & 0.970 & 0.113 & 0.161 & 22.445 & 0.150 \\
  RCV-LASSO & -0.029 & 0.147 & 33.470 & 0.990 & 0.046 & 0.161 & 31.890 & 0.450 \\
  \hline
  P-SCAD & -0.036 & 0.096 & 6.110 & 1.000 & -0.066 & 0.102 & 14.520 & 1.000 \\
  CV-SCAD & 0.003 & 0.096 & 6.110 & 1.000 & 0.079 & 0.124 & 14.520 & 1.000 \\
  P-LASSO & -0.097 & 0.171 & 39.290 & 1.000 & -0.089 & 0.171 & 37.140 & 1.000 \\
  CV-LASSO & 0.126 & 0.116 & 39.290 & 1.000 & 0.125 & 0.116 & 37.140 & 1.000 \\
   \hline\hline
\end{tabular}

\end{table}

To show the effectiveness of $\hat{\sigma}_{\sRCV}$ in the construction of confidence intervals, we calculate the coverage probability of the confidence interval ({\ref{eqci}) based on 10,000 simulations. This was conducted for $\beta_1$, $\beta_2$ and $\beta_3$ with $b = 1/\sqrt{3}$, 1, and 2 and $\rho = 0$ and $0.5$.  To save the space of the presentation, we present only one specific case for $\beta_1$ with $b=1$ in Table 3.

\begin{table}[htbp]\scriptsize
\caption{Simulation results for Example 2 with $n=200$, $p=2000$, $b=1$: coverage probability of confidence intervals of different levels for $\beta_1$, based on 10000 replications.  }
\centering
\begin{tabular}{|r|rrrr|rrrr|}
  %\hline
  %\multicolumn{9}{c}{$ b=1$} \\
  \hline
  &     & $\rho=0$&  &      &      & $\rho=0.5$ &     &    \\
  \hline
 & 80\% & 90\% & 95\% & 99\% & 80\% & 90\% & 95\% & 99\%    \\
  \hline
Oracle & 0.7967 & 0.8974 & 0.9476 & 0.9874 & 0.7931 & 0.9006 & 0.9483 & 0.9865 \\
RCV & 0.7919 & 0.8928 & 0.9435 & 0.9847 & 0.8042 & 0.9022 & 0.9518 & 0.9871\\
   \hline    \hline
\end{tabular}

\end{table}

%%%%%%%%%%%%%%%%%%%%%%%%%%%%%%%%%%%%%%%%%%%%%%%%%%%%%%%%%%%%%%%%%%%%%%%%%%%%%%
%
%                                Example 3
%
%%%%%%%%%%%%%%%%%%%%%%%%%%%%%%%%%%%%%%%%%%%%%%%%%%%%%%%%%%%%%%%%%%%%%%%%%%%%%%%
{\bf Example 3.} We consider a more realistic model with 10 important predictors, detailed at beginning of this section. Since some non-vanishing coefficients are very small, no method can guarantee all relevant variables are chosen in the selected model, i.e. possess a sure screening property. To quantify the severity of missing relevant variables, we use the quantity Variance of Missing Variables (VMV), $\var(\bx_S^T \bbeta_S)/\sigma^2$ to measure, where $S$ is the set of important variables not included in the selected model and $\bbeta_S$ is their regression coefficients in the simulated model. For RCV methods, the VMV is the average of VMVs for two halves of the data. Figure \ref{fig9} summarizes the simulation results for $(n,p)=(400,1000)$, whereas Figure \ref{fig10} depicts the results for $(n,p)=(400,10000)$ when the penalization methods are not easily accessible.  The naive methods seriously underestimate the variance and sensitive to the model selection tools, dimensionality, SNR, among others. In contrast, the RCV methods are much more stable and only slightly overestimate the variance when the sure screening condition is not satisfied.  The one-step methods, especially plug in methods, perform also well.

\begin{figure}[htbp]
\includegraphics[width=80mm,height=162mm,angle=270]{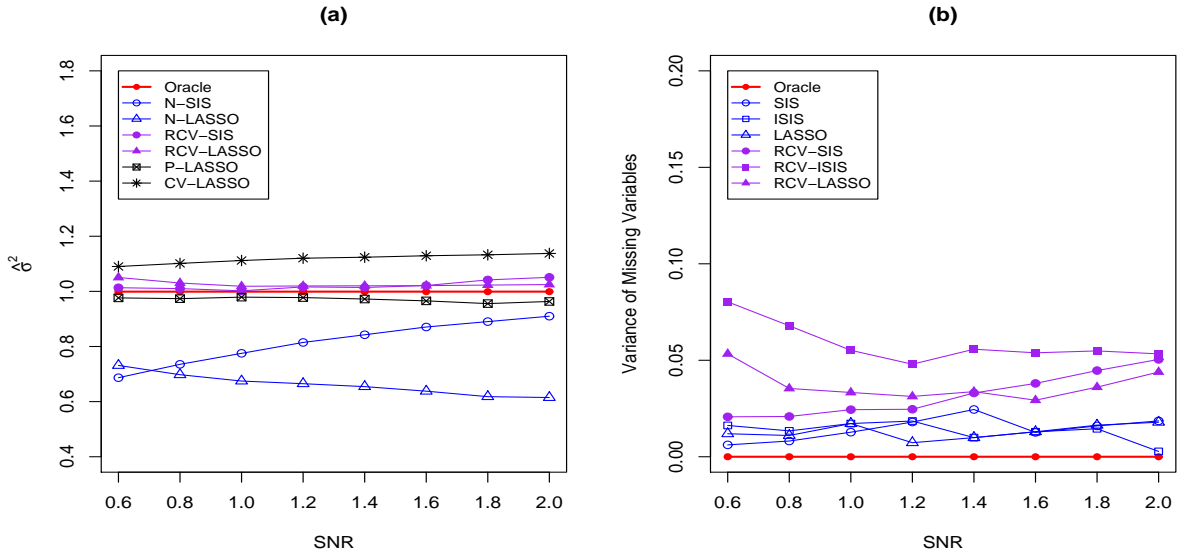}
\caption{(a) The medians of various variance estimators when $n=400$ and $p=1000$ among $100$ replications for Example 3.  (b) The medians of variance of missing variables of different model selection methods.}\label{fig9}
\end{figure}
\begin{figure}[htbp]
\includegraphics[width=80mm,height=162mm,angle=270]{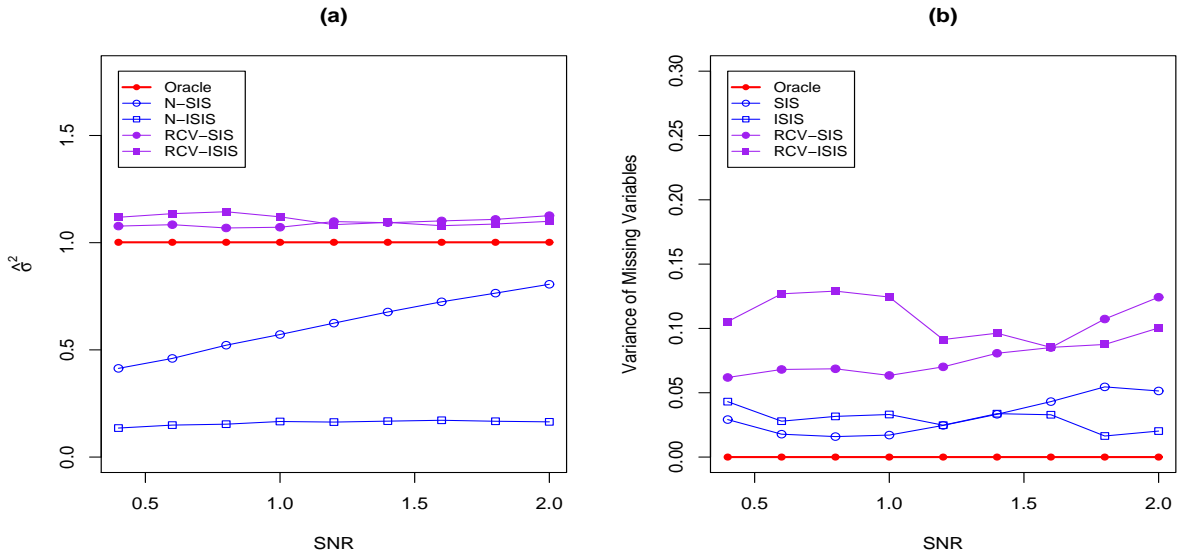}
\caption{(a) The medians of various variance estimators when $n=400$ and $p=10000$ among $100$ replications. (b) The medians of variance of missing variables of different model selection tools. }\label{fig10}
\end{figure}

%%%%%%%%%%%%%%%%%%%%%%%%%%%%%%%%%%%%%%%%%%%%%%%%%%%%%%%%%%%%%%%%%%%%%%%%%%%%%%%%%%%%%%%
%
%                     Real data Analysis
%
%%%%%%%%%%%%%%%%%%%%%%%%%%%%%%%%%%%%%%%%%%%%%%%%%%%%%%%%%%%%%%%%%%%%%%%%%%%%%%%%%%%%%%%

\subsection{Real data analysis}

We now apply our proposed procedure to analyze a recent house price data from 1996-2005. The data set consists of 119 months of appreciations of national House Price Index (HPI), defined as the percent of monthly log-HPI changes in 381 Core Based Statistical Areas (CBSA) in the United States. The goal is to forecast the housing price appreciation (HPA) over those 381 CBSAs over the next several years.  Housing prices are geographically dependent.  They depend also on macroeconomic variables.  Their dependence on macroeconomic variables can be summarized by the national HPA.  Therefore, a reasonable model for predicting the next period HPA in a given CBSA is
\begin{equation} \label{a14a}
   Y_t = \beta_0 + \beta_N X_{t-1, N} + \sum_{i = 1}^{381} \beta_{i} X_{t-1, i} + \varepsilon_t,
\end{equation}
where $X_N$ stands for the national HPA, $\{X_{i}\}_{i=1}^{381}$ are the HPAs in those 381 CBSAs, and $\varepsilon$ is a random error independent of $X$. This is clearly a problem with the number of predictors more than the number of covariates. However, conditional on the national HPA $X_N$, it is reasonable to expect that only the local neighborhoods have non-negligible influence, but it is hard to pre-determine those neighborhoods. In other words, it is reasonable to expect that the coefficients $\{\beta_i\}_{i=1}^{381}$ are sparse.

Our primary interest is to estimate the residual variance $\sigma^2$, which is the prediction error of the benchmark model.  We always keep the variables $X_N$ and $X_1$, which is the lag 1 HPA of the region to be predicted.  We applied the SCAD using the local linear approximation \citep{ZouLi08}, which is the iteratively re-weighted LASSO, to estimate coefficients in (\ref{a14a}).  We summarize the result, $\hat{\sigma}$, as a function of the selected model size $s$, to examine the sensitivity to the selected model size.  Reported also is the percent of variance explained which is defined as
$$
    R^2 = 1 - \frac{\mbox{RSS}}{ \sum_{t=1}^{119} (Y_t - \bar{Y})^2},
$$
where $\bar{Y}$ is the sample average of the time series.  For illustration purpose, we only focus on one CBSA in San Francisco and one in Los Angeles.  The results are summarized in Table 3 and Figure \ref{figreal}, in which the naive two-stage method is also included for comparison.

First of all, as shown in Figure \ref{figreal}, the influence of the naive method by the selected model size is much larger than that of the RCV method.  This is due to the spurious correlation as we discussed before.  The RCV estimate is reasonably stable, but it is also influenced by the selected model size  when it is large.  This is understandable given the sample size of 119.

In the case of San Francisco, from Figure \ref{figreal}(b), the RCV method suggests that the standard deviation should be around $0.52\%$, which is reasonably stable for $s$ in the range of $4$ to 8. By inspection of the solution path of the naive two-stage method, we see that besides $X_N$ and $X_1$, first selected is the variable $X_{306}$, which corresponds to CBSA San Jose-Sunnyvale-Santa Clara (San Benito County, Santa Clara County). The variable $X_{306}$ also enters into both models when $s\geq 3$ in the RCV method. Therefore, we suggest that the selected model consist of at least variables $X_1$, $X_2$ and $X_{306}$. As expected, in the RCV method, the fourth selected variables are not the same for the two splitted subsamples. The variance explained by regression takes $79.83\%$ of total variance.

Similar analysis can be applied to the Los Angeles case. Figure \ref{figreal}(d) suggests the standard deviation should be around $0.50\%$ (when $s$ is between $7$ and $10$). From the solution path,
we suggest that the selected model consist of at least variables $X_N$, $X_1$ and $X_{252}$ which corresponds to CBSA Oxnard-Thousand Oaks-Ventura (Ventura County).
The variance explained by regression takes $90.23\%$ of total variance.

%To see whether it is statistically significant, we calculated the estimated standard error of the variance,

%If we do regression with first three rows, the estimated $\hat{\sigma} = 0.2739\%$.

%\end{document}

%%%%%%%%%%%%%%%%%%%%%%%%%%%%%%%%%%%%%%%%%%%%%%%%%%%%%%%%%%%%%%%%%%%%%%%%%%%%%%%%%%%%%%%%%%%%%%%%%%%%%%%%%%%%
\begin{table}[htbp]\scriptsize 
\caption{\small Estimated residual standard deviation and  variance explained by regression (in percent) for naive two-stage and RCV methods for forecasting home price appreciation in San Francisco and Los Angeles.}\label{realdata}

\centering
 \begin{tabular}{|l|ccccccc|}
 \multicolumn{8}{c}{San Francisco}\\
 \hline \hline
  Model size & 2 & 3 & 5 &10 & 15 & 20 & 30 \\
  \hline
  Naive       & 0.5577  & 0.5236 & 0.5072& 0.4555 & 0.3938 &0.3862 & 0.3635\\
  \hline
  RCV  & 0.5563 &  0.5536& 0.5179 & 0.5057 & 0.4730 & 0.4749& 0.4735\\
    \hline
  variance explained & 76.92 &  79.83& 81.40 & 85.67 & 89.79 & 90.66& 92.58\\
   \hline \hline
   \multicolumn{8}{c}{Los Angeles}\\
 \hline \hline
  Model size & 2 & 3 & 5 &10 & 15 & 20 & 30 \\
  \hline
  Naive       &  0.5236  & 0.4887 & 0.4583& 0.4401 & 0.3747 &0.3137 & 0.2503\\
  \hline
  RCV     & 0.5255 &  0.5214 & 0.5210& 0.4995 & 0.4794 & 0.4596 & 0.4621\\
    \hline
  variance explained & 88.68 &  90.23& 91.56 & 92.56 & 94.86 & 96.57& 98.05\\
   \hline \hline
 \end{tabular}
\end{table}

\begin{figure}[htbp]
\centerline{
\includegraphics[width = 150mm, height = 150mm]{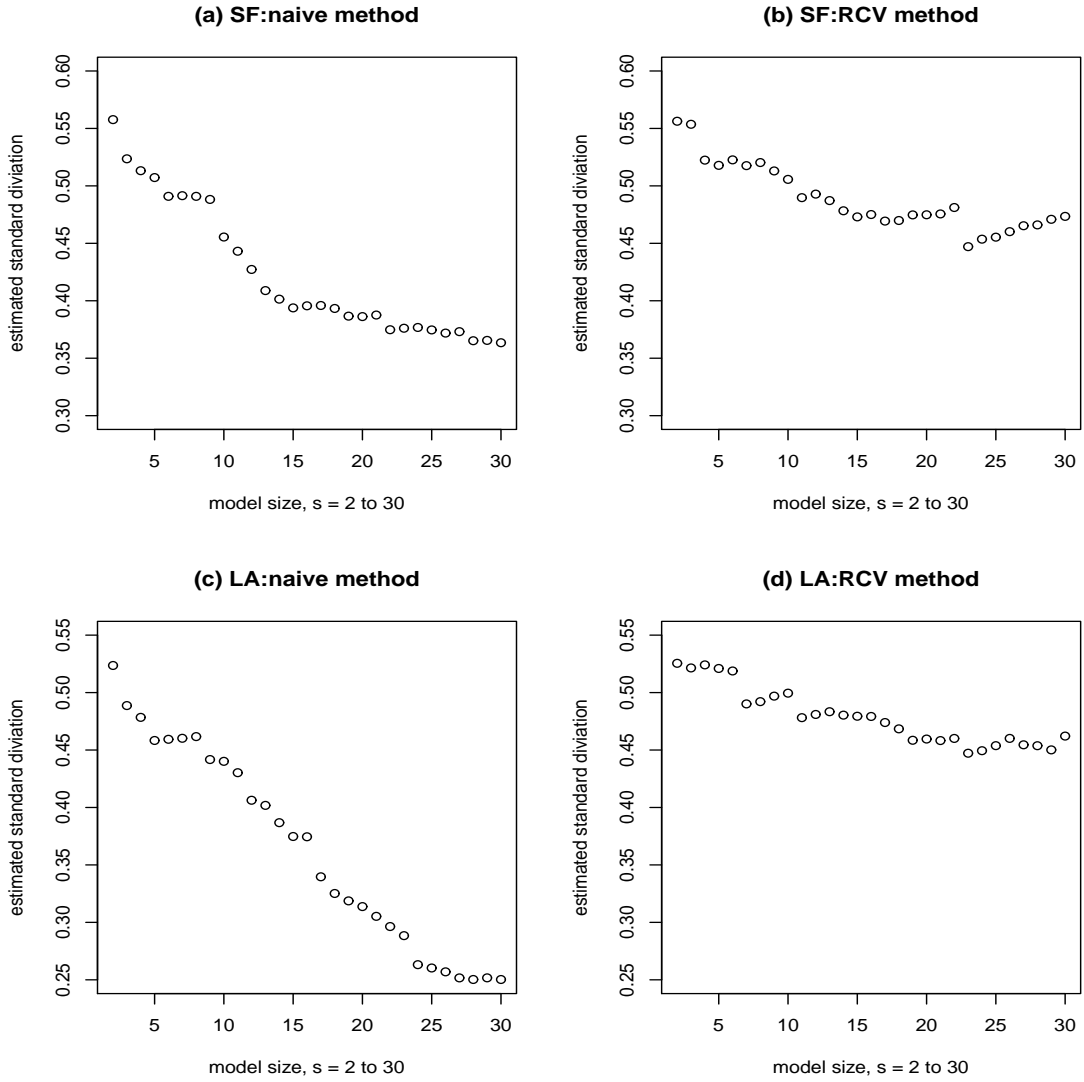}}
\caption{Estimated standard deviation of benchmark one-step forecast of home price appreciation in San Francisco and Los Angeles for various selected model size. The results are based on both the naive two-stage and RCV methods.}\label{figreal}
\end{figure}

%%%%%%%%%%%%%%%%%%%%%%%%%%%%%%%%%%%%%%%%%%%%%%%%%%%%%%%%%%%%%%%%%%%%%%%%%%%%%%%%%%%%%%%
%\clearpage
\section{Discussion}\label{sec:dis}
Variance estimation is important and challenging for ultrahigh dimensional sparse regression.  One of the challenges is the spurious correlation:  covariates can have high correlations with the realized noise and hence are recruited to predict the noise.  As a result, the naive (natural) two-stage estimator seriously underestimates the variance. Its performance is very unstable and depends largely on the model selection tool employed. The RCV method is proposed to attenuate the influence of the effect of spurious variables. Both the asymptotic theory and empirical result show that the RCV estimator is the best among all estimators.  It is accurate and stable, insensitive to the model selection tool and the size of the selected model. Therefore, we may employ fast model selection tool like SIS for computational efficiency for the RCV variance estimation. We also compare the RCV method with the direct plug-in method. When choosing tuning parameters of a penalized likelihood method like the LASSO, we suggest using a more conservative cross-validation rather than aggressive BIC. However, the LASSO method can still yield a non-negligible bias for variance estimation in ultrahigh dimensional regression. The SCAD method is almost as good as the RCV method, but it is computational more expensive than RCV-SIS.

\appendix
\section*{Appendix}
{\bf Notation and conditions}

\vspace{2ex}

We first state the following assumptions, which are standard in the literatures of high dimensional statistical learning.
For convenience, define
$
 \phi_{\min}(m) = \min_{M: |M| \le m} \lambda_{\min}({1 \over n}{\mathbf{X}}_M^T {\mathbf{X}}_M)
$
 and
$
\phi_{\max}(m) = \max_{M: |M| \le m} \lambda_{\max}({1 \over n}{\mathbf{X}}_M^T {\mathbf{X}}_M),
$
where $\lambda_{\min}(\mathbf{A})$ and $\lambda_{\max}(\mathbf{A})$ denote the smallest and largest eigenvalues of a matrix $\mathbf{A}$, respectively.

For a vector $\mathbf{v}$, we use standard natation $||\mathbf{v}||_p=\left(\sum_i |v_i|^p\right)^{\frac1p}$ and $||\mathbf{v}||_{\infty}=\max_i\{|v_i|\}$. For a matrix $\mathbf{B}$, we use three different norms. $||\mathbf{B}||_{2,\infty}$ is defined in Assumption (A8) below; $||\mathbf{B}||_{2}$ denotes the usual operator norm, i.e. $||\mathbf{B}||_{2 } = \max_{||\bs{v}||_2 \le 1}||\mathbf{B}\bs{v}||_2$; $||\mathbf{B}||_{\infty}=\max_{i,j}\{|B_{ij}|\}$ is the usual sup-norm.
 \vspace{3ex}

\begin{description}
\item[(A1)]
\label{assum:err}
The errors $\varepsilon_1,...,\varepsilon_n$ are $i.i.d.$ with zero mean and finite variance $\sigma^2$ and independent of the design matrix $\mathbf{X}$.
\item[(A2)]
\label{assum:eig}
 There exists a constant $\lambda_0>0$ and $b_n$ such that $b_n / n \to 0$ such that $ P\Big\{\phi_{\min}(b_n) \ge \lambda_0\Big\}=1$
 for all $n$.
\item[(A3)]
\label{assum:rand}
There exists a constant $L$ such that $\max_{i,j}|X_{ij}| \le L$, where $X_{ij}$ is the $(i, j)$ element of the design matrix $\bX$.

\item[(A4)\label{assum:sub}]
$\E\{\exp(|\varepsilon_1|/a)\}\le b$ for some finite constants $a,b>0$.
%\end{assumption}
\end{description}

We have no intent to make the assumptions the weakest possible. For example, Assumption (A3) can be relaxed to $\max_{i,j} |X_{ij}| \le L (\log n)^\xi$ for any $\xi>0$ or further relaxation.  The aim  of the assumptions (A3) and (A4) is to guarantee that $\hat{\gamma}_n$ in Theorem 1 is of the order $\sqrt{\hat{s} \log p/n}$.

Theorem 1 still holds under the random design with assumptions below.

\begin{description}
\item[(A5)] \label{assum:rand2}
The random vectors $\mathbf{x}_{1},..., \mathbf{x}_{n}$ are $i.i.d.$ and there exists a constant $\alpha$ such that $\mbox{E}\big[\exp\{(|X_{ij}|/\rho)^\alpha\}\big] \le L$ for all $i,j$ and some constants $\alpha>1$, and $\rho, L >0$, where $X_{ij}$ is the $(i,j)$th element of $\mathbf{X}$.

\item[(A6)]\label{assum:sub2} $\varepsilon_1$ satisfies that
$ \mbox{E}\big[\exp\{(|\varepsilon_1|/a)^\theta\}\big]\le b$ for some finite positive constants $a,b,\theta >0$ and $1/\alpha + 1/\theta \le 1$, where $\alpha$ is defined by Assumption (A5).
\end{description}
For instance, when $X_{ij}$ and $\varepsilon_i$ are sub-Gaussian ($\alpha=\theta=2$) for each $i$ and $j$, the assumptions (A5) and (A6) are satisfied.

The following assumption (A7) is imposed for proving Theorem 3. For fixed design matrix $\mathbf{X}$, the corresponding condition is also imposed in \citet{mein:yu:lass:2009} and some discussions of weaker conditions are shown in \citet{bick:rito:simu:2008}.

\begin{description}
\item[(A7)] \label{assum:eig2}
 There exist constants $0 < k_{\min} \le k_{\max} < \infty$  such that
 $$
 P\big\{\liminf_{n \rightarrow \infty} \phi_{\min}(s \log n) \ge k_{\min}\big\} = 1,
 $$
 and \hskip 2cm
 $$
 P\big\{\limsup_{n \rightarrow \infty} \phi_{\max}(s + \min\{n,p\}) \le k_{\max}\big\} = 1.
 $$
 \end{description}

The following two additional assumptions are stated for proving Theorem 4. These conditions correspond to Conditions 4 and 5 in \citet{Fan:Lv:NP:2009}. Without loss of generality, assume that the true value $\bbeta_0 = (\bbeta_{01}^T, \bbeta_{02}^T)^T$ with each component of $\bbeta_{01}$ nonzero and $\bbeta_{02} = \bs{0}$. Let $\mathbf{X}_1$ and $\mathbf{X}_2$ be the submatrices of $n \times p$ design matrix  ${\mathbf{X}}$ with columns corresponding to $\bbeta_{01}$ and $\bbeta_{02}$, respectively.

 \begin{description}
 \item[(A8)]\label{assum:eig3}
 There exist constants $0 < c_1,c_2 < \infty$  such that
 $$
 P\Big\{ \lambda_{\min}\big({1 \over n}{\mathbf{X}}_1^T {\mathbf{X}}_1\big)\ge c_1\Big\} \rightarrow 1,
 $$
 and
 $$
 P\Big\{||{1 \over n}{\mathbf{X}}_2^T {\mathbf{X}}_1||_{2,\infty} \le c_2\Big\} \rightarrow 1,
 $$
 as $n \rightarrow \infty$, where
 $
 ||\mathbf{B}||_{2,\infty} = \max_{||\bs{v}||_2 \le 1}||\mathbf{B}\bs{v}||_\infty.
 $
\item[(A9)] \label{assum:lambda}
 Denote $d_n = {1 \over 2} \min_{j =1, \cdots,s} |\beta_{0j}|.$
 Assume that $d_n \ge n^{-\gamma} \log n$ with $\gamma \in (0,1/2]$. Take $\lambda_n \propto n^{- {1-\alpha_0 \over 2}}\log n$ and $\lambda_n \ll d_n$, where $\alpha_0$ is defined in Theorem \ref{thm4}.
 \end{description}

{\bf Remark}:  The norm $\| \bB \|_{2, \infty}$ is somewhat abstract.  It can easily be shown that
$$
   \| \bB \|_{2, \infty} \leq s \| \bB \|_\infty,
$$
where $s$ is the number of columns of $\bB$, which is a crude upper bound.  Using this and the argument in the proof of Theorem 4, if
$$
   P\Big\{||{1 \over n}{\mathbf{X}}_2^T {\mathbf{X}}_1||_{\infty} \le c_3 \Big\} \rightarrow 1
$$
and $\lambda_n \ge n^{-(1-3 \alpha_0)/2} \log n$ and $\lambda_n \ll d_n$, then the conclusion of Theorem~\ref{thm4} holds.

 {\bf A.1. Proof of Theorem~\ref{thm1}}

\vspace{2ex}

Part 1 follows the standard law of large numbers and central limit theorem.
Now we prove the second part under assumptions (A1)- (A4).
%
%Denote $\mathcal{A}_{n}=\{M_0 \subset \hat{M}\}$. On the event $\mathcal{A}_n$, we have
%%By assumptions (\ref{assum:sis}) and (\ref{assum:spar}), with a probability close to 1,
%\[
%\hat{\sigma}_n^2 = \frac{\boldsymbol\varepsilon^T(\mathbf{I}_n-\mathbf{P}_{\hat M})\boldsymbol\varepsilon}{n-\hat{s}}.
%\]
By Assumption (A2),
\begin{equation}
\beps^T\mathbf{P}_{\hat M}\beps=
\beps^T\mathbf{X}_{\hat M}(\mathbf{X}_{\hat M}^T\mathbf{X}_{\hat M})^{-1}\mathbf{X}_{\hat M}^T\beps \leq \frac{1}{\lambda_0n} \pp\mathbf{X}_{\hat M}^T\beps \pp^2.\label{eqn:a23}
\end{equation}
Let $\bX_{j}$ denote the $j$-th column vector of the design matrix $\mathbf{X}$.
%The next part of the proof is following the proof of Lemma 5 in \cite{bune:tsyb:spar:2007}. For the completeness, we cited the proof here.
For a large constant $c$, consider the event
$
\mathcal{E}_n= \left\{\max_{1 \leq j \leq p} |\bX_{j}^T\boldsymbol{\varepsilon}| \leq c\sqrt{n\log p}\right\}.
$
Under the event $\mathcal{E}_n$, it follows from equation (\ref{eqn:a23}) that
\begin{align*}
\beps^T\mathbf{P}_{\hat M}\beps  \leq {1 \over \lambda_0}\hat s c^2\log p.
\end{align*}
Together with the fact $n^{-1} \|\beps\|^2 \to \sigma^2$,  we get
$$
\hat{\gamma}_n^2 =\beps^T\mathbf{P}_{\hat M}\beps/\beps^T\beps= O_P(\hat{s}\log p /n).$$
Hence it suffices to show that $P(\mathcal{E}_n) \to 1$ as $n\to \infty$ for some constant $c$.
Observe that, by Assumptions (A3)-(A4), for each $j$,
$$
\E |X_{ij}\varepsilon_i|^m \le m!(La)^m\E\exp\{|\varepsilon_1|/a\}\le {1 \over 2}m!(2ba^2L^2)(aL)^{m-2}.
$$
Using Bernstein's inequality (e.g. Lemma 2.2.11 of \citet{VW96} ), we have
\begin{eqnarray}
\label{T1-1}
P\{\mathcal{E}_n^C\} &\le& P\Big\{\max_{1 \le j \le p}|\bX_{j}^T\bs{\varepsilon}| \ge c\sqrt{n\log p}\Big\}\nonumber\\
 &\le& \sum_{j=1}^p P\Big\{|\bX_j^T\bs{\varepsilon}| \ge c\sqrt{n\log p}\Big\}\nonumber\\
&\le& 2p \cdot \exp\Big\{-{c^2n\log p \over 2(2ba^2L^2 + aL\cdot c\sqrt{n\log p})}\Big\}\nonumber\\
\label{a16}
&=& 2\exp\Big\{ \log p\Big(1-{1 \over 4ba^2L^2c^{-2}n^{-1}+2aLc^{-1}\sqrt{\log p/n}}\Big) \Big\}
\end{eqnarray}

For sufficient large $c$, we have $4ba^2L^2c^{-2}n^{-1}+2aLc^{-1}\sqrt{\log p/n}<1$ since $\log p /n$ is bounded. Therefore, the power in (\ref{a16}) goes to negative infinity as $p\rightarrow \infty$. It follows that $P\{\mathcal{E}_n\} =  1- P\{\mathcal{E}_n^C\} \rightarrow 1$.

Next we show the second part of the theorem still holds under Assumptions (A5)-(A6) instead of Assumptions (A3)-(A4).
It is sufficient to verify $P(\mathcal{E}_n) \to 1$ as $n\to \infty$ for some constant $c$. The key step is to establish the inequality
\begin{eqnarray}
\label{T1-4}
E\{|X_{ij}\varepsilon_i|^m\} \le {1 \over 2}m!\Big(8(2+ L + b) \rho^2 a^2\Big) (2\rho a)^{m-2},
\end{eqnarray}
for each $j=1,\cdots,p$.
%then, by using Bernstein's equality with the similar arguments as (\ref{T1-1}), we have that
%$$
%P\{\mathcal{E}_n^C\} \rightarrow 0,
%$$
%and thus, $\hat{\gamma}_n =\sqrt{\hat s \log p/n}$. It suffices to prove (\ref{T1-4}).
Note that
$$
P\{|XY|> t\} \le P\{|X|>t^{1/\alpha}\}+ P\{|Y|>t^{1-1/\alpha}\}
$$
for $\alpha>1$ and random variables $X$ and $Y$. Thus,  for any $t \ge 1$ and each $i,j$,
\begin{eqnarray*}
P\Big\{\Big|{X_{ij}\over \rho}\Big|\cdot\Big|{\varepsilon_i \over a}\Big| > t\Big\}
&\le& P\Big\{\Big|{X_{ij}\over \rho}\Big| > t^{1 /\alpha}\Big\}+P\Big\{\Big|{\varepsilon_i \over a}\Big| > t^{1 - {1/\alpha}}\Big\}\\
&\le& L \exp \{-t\} + b \exp\{-t^{\theta(1 -1/\alpha)}\}\\
&\le& (L + b) \exp(-t).
\end{eqnarray*}
If $X$ is a nonnegative random variable with its distribution $F(t)$ and tail probability $P\{X>t\}\leq C\exp(-t)$ for some constant $C > 0$ and each $t \ge 1$, then by integration by parts
\begin{eqnarray*}
\E \exp\big(\frac 12 X\big) &=& -\int_0^{\infty}\exp\big(\frac x2\big) d \big\{1 - F(x)\big\}\\
                &=& 1 + \frac12\int_0^{\infty}\{1-F(x)\}\exp\big(\frac x2\big)dx \\
               % && \hskip 3cm \mbox{(\bf using integration by parts)}\\
                &\leq& 1 +\frac12\int_0^1 \exp\big(\frac x2\big)dx + \frac12\int_1^{\infty}C\exp\big(- \frac x2\big)dx\\
                &\leq& 2 + C.
\end{eqnarray*}
As a result, it follows that, for each $i,j$,
\begin{eqnarray*}
\E\exp\Big\{{1\over 2}\Big|{X_{ij}\over \rho}\Big|\cdot\Big|{\varepsilon_i \over a}\Big|\Big\}
\le  2+ (L + b).
\end{eqnarray*}
%As a result, it follows that, for each $i,j$,
%\begin{eqnarray*}
%\E\exp\Big\{{1\over 2}\Big|{X_{ij}\over \rho}\Big|\cdot\Big|{\varepsilon_i \over a}\Big|\Big\}
%&\le& {1 \over 2}\int_0^\infty\exp\big({x\over 2}\big)\cdot \max\{L,b\}\exp(-x)dx+
%{1 \over 2}\int_0^1\exp\big({x\over 2}\big)dx\\
%&\le&1+\max\{L,b\}.
%\end{eqnarray*}
Thus, for each positive integer $j$ and $m \ge 2$,
\begin{eqnarray*}
\E\{|X_{ij}\varepsilon_i|^m\}
&\le & (2\rho a)^m m!\E\exp\Big\{{1\over 2}\Big|{X_{ij}\over \rho}\Big|\cdot\Big|{\varepsilon_i \over a}\Big|\Big\}\\
&\le&(2\rho a)^m m! (2+ L + b ) \\
&=&{1 \over 2}m!\Big( 8(2+ L + b) \rho^2 a^2\Big)(2\rho a)^{m-2}.
\end{eqnarray*}
Theorem 1 is proved.

 \vspace{3ex}

{\bf A.2. Proof of Theorem~\ref{thm2}}

\vspace{2ex}
Define sequences of events $\mathcal{A}_{n1}=\{M_0 \subset \hat{M}_1\}$, $\mathcal{A}_{n2}=\{M_0 \subset \hat{M}_2\}$ and $\mathcal{A}_n=\mathcal{A}_{n1}\cap \mathcal{A}_{n2}$.
%By assumptions (A5) and (A6), with probability close to 1,
On the event $\mathcal{A}_n$, we have
\begin{align*}
\hat\sigma_1^2  = \frac{({\beps}^{(2)})^T(\mathbf{I}_{n/2} - \mathbf{P}^{(2)}_{\hat M_1}){\beps}^{(2)}}{{n/2}-\hat{s}_1}\hskip 0.5cm \mbox{and} \hskip 0.5cm \hat\sigma_2^2 = \frac{({\beps}^{(1)})^T(\mathbf{I}_{n/2}-{\mathbf{P}}^{(1)}_{\hat M_2}){\beps}^{(1)}}{{n/2}-\hat{s}_2},
\end{align*}
where ${\beps}^{(1)}$ and ${\beps}^{(2)}$ correspond to $\mathbf{y}^{(1)}$ and $\mathbf{y}^{(2)}$, respectively. Decompose now  $(n/2-\hat{s}_1)(\hat\sigma_1^2-\sigma^2)$ on the event $\mathcal{A}_n$ as
\begin{align*}
 ({1\over 2}n-\hat{s}_1)(\hat\sigma_1^2-\sigma^2)=
 \big\{({\beps}^{(2)})^T{\beps}^{(2)} - {1\over 2}n\sigma^2 \big\}-
 \big\{({\beps}^{(2)})^T\mathbf{P}^{(2)}_{\hat M_1}{\beps}^{(2)}-
 \hat{s}_1\sigma^2\big\}.
 \end{align*}
We now prove $({\beps}^{(2)})^T\mathbf{P}^{(2)}_{\hat M_1}{\beps}^{(2)}- \hat{s}_1\sigma^2 = O_P(\sqrt{\hat{s}_1})$.

First, consider the quadratic form $S = \bs{\xi}^T \mathbf{P}\bs{\xi}$ where $\mathbf{P}$ is a symmetric $m \times m $ matrix, $\bs{\xi} = (\xi_1,\cdots,\xi_m)^T$ and $\xi_i$ $(i=1,\cdots,m)$ are $i.i.d$. Assume that $E\xi_1 = 0$, $\E\xi_1^2 = \sigma^2$ and the fourth moment $\E\xi_1^4 <\infty$. Let $P_{ij}$ be the $(i,j)$th element of the matrix $\mathbf{P}$. Then,
\begin{eqnarray*}
\E(S) = \E\big(\sum_{i=1}^m \xi_i^2P_{ii}\big) = \sigma^2 \cdot \hskip 0.1cm trace(\mathbf{P}),
\end{eqnarray*}
and
\begin{eqnarray*}
\Var{(S)} %&=& \E\big(\sum_{i=1}^m \xi_i^2P_{ii}\big)^2 - \big(\E(S_n)\big)^2  \\
 &=& \E\big(\sum_{i,j,l,k}^m\xi_i\xi_j\xi_l\xi_kP_{ij}P_{lk}\big) - \sigma^4 \cdot (\sum_{i=1}^m P_{ii})^2\\
 &=& \E\big(\sum_{i=1}^m\xi_i^4P_{ii}^2\big) + \E\big(\sum_{i=l\neq j=k}^m\xi_i^2\xi_j^2P_{ij}P_{lk}\big)  + \hskip 0.2cm
 \E\big(\sum_{i=k\neq j=l}^m\xi_i^2\xi_j^2P_{ij}P_{lk}\big)\\
 && + \E\big(\sum_{i=j\neq l=k}^m\xi_i^2\xi_l^2P_{ij}P_{lk}\big) - \sigma^4 \cdot (\sum_{i=1}^m P_{ii})^2\\
 &=& \E\xi_1^4\cdot \big(\sum_{i=1}^mP_{ii}^2\big) + 2\sigma^4\cdot\big(\sum_{i\neq j}^mP_{ij}^2\big) + \sigma^4\cdot\big(\sum_{i\neq l}^mP_{ii}P_{ll}\big) - \sigma^4 \cdot (\sum_{i=1}^m P_{ii})^2\\
% & = & (\E\xi_1^4 - \sigma^4)\big(\sum_{i=1}^mP_{ii}^2\big)+ 2\sigma^4\cdot\big(\sum_{i\neq j}^mP_{ij}^2\big) + \sigma^4\cdot\big(\sum_{i,l}^mP_{ii}P_{ll}\big)   - \sigma^4 \cdot (\sum_{i=1}^m P_{ii})^2\\
 %& = & (\E\xi_1^4 - \sigma^4)\big(\sum_{i=1}^mP_{ii}^2\big)+ 2\sigma^4\cdot\big(\sum_{i\neq j}^mP_{ij}^2\big)\\
 & = & (\E\xi_1^4 - \sigma^4)\big(\sum_{i=1}^mP_{ii}^2\big)+ 2\sigma^4\cdot\big(\sum_{i\neq j}^mP_{ij}^2\big) \\
 & \le & (\E\xi_1^4 + \sigma^4)\cdot trace\big(\mathbf{P}^2\big).
\end{eqnarray*}
where the last inequality holds since $trace\big(\mathbf{P}^2\big) = \sum_{i,j}^m P_{ij}^2$.

Observe that, $trace\big(\mathbf{P}^{(2)}_{\hat M_1}\big) = trace\big\{(\mathbf{P}^{(2)}_{\hat M_1})^2\big\} = \hat{s}_1$. Hence, on the event $\mathcal{A}_{n1}$, we have
$$
\E\Big\{({\beps}^{(2)})^T\mathbf{P}^{(2)}_{\hat M_1}{\beps}^{(2)}
 \Big|\bX^{(2)}_{\hat M_1}\Big\}=\hat{s}_1\sigma^2,
$$
and
$$
 \Var\Big\{({\beps}^{(2)})^T \mathbf{P}^{(2)}_{\hat M_1}{\beps}^{(2)}\Big|\bX^{(2)}_{\hat M_1}\Big\}
 \le (E\varepsilon^4+\sigma^4)\hat{s}_1.
$$

Using Markov inequality, it follows that, under the event $\mathcal{A}_{n1}$,
 \begin{eqnarray*}
 ({\beps}^{(2)})^T\mathbf{P}^{(2)}_{\hat M_1}{\beps}^{(2)}-
 \hat{s}_1\sigma^2  & = & O_P(\sqrt{\hat{s}_1}).
 \end{eqnarray*}
 Combining with the assumptions $\hat{s}_1/n \stackrel{P}{\rightarrow}0 $ and $P(\mathcal{A}_{n1})\stackrel{P}{\rightarrow}1$, we obtain that
$$
 ({\beps}^{(2)})^T\mathbf{P}^{(2)}_{\hat M_1}{\beps}^{(2)}-
 \hat{s}_1\sigma^2=o_P(\sqrt{n}).
$$
 As a result,
 $$
 ({1\over 2}n-\hat{s}_1)(\hat\sigma_1^2-\sigma^2)=
 ({\beps}^{(2)})^T{\beps}^{(2)}-{1\over 2}n\sigma^2 + o_P(\sqrt{n}).
 $$
 Similarly, we conclude that
 $$
 ({1\over 2}n-\hat{s}_2)(\hat\sigma_2^2-\sigma^2)=
 ({\beps}^{(1)})^T{\beps}^{(1)}-{1\over 2}n\sigma^2 + o_P(\sqrt{n}).
 $$
Therefore, using the last two results, we have
 \begin{align*}
 &\sqrt{n}(\hat{\sigma}_{\sRCV}^2-\sigma^2)\\
 &={\sqrt{n} \over {n-2\hat{s}_1}}\Big\{ ({\beps}^{(2)})^T{\beps}^{(2)}-{1\over 2}n\sigma^2\Big \}
 + {\sqrt{n} \over {n-2\hat{s}_2}}\Big\{ ({\beps}^{(1)})^T {\beps}^{(1)}-{1\over 2}n\sigma^2 \Big \}+o_P(1)\\
  & ={1 \over \sqrt{n}}\sum_{i=1}^n(\varepsilon_i^2-\sigma^2)+o_P(1),
 \end{align*}
 which implies that $\sqrt{n}(\hat{\sigma}_{\sRCV}^2-\sigma^2)\stackrel{\mathcal{D}}{\rightarrow} N(0,\E[\varepsilon^4]-\sigma^4)$.
 The Proof of  Theorem 2 is completed.

 \vspace{2ex}

To prove Theorem 3, we will use the following lemma. The results are stated and proved in \citet{mein:yu:lass:2009} and \citet{bick:rito:simu:2008}.
% for fixed design matrix ${\mathbf{X}}$, but their conditions holds with probability tending to one and so the results still hold for random matrix ${\mathbf{X}}$.

\vspace{2ex}

\begin{lemma}   \label{lem1}  %Lemma 1
Consider the LASSO selector $\hat{\bs \beta}_{L}$ defined by (\ref{a8}) with $\lambda_n $.  Under the assumptions (A1)-(A4) and (A7),
for $\lambda_n \propto \sigma\sqrt{\log p/n}$, there exists a constant $M>0$ such that, with probability tending to 1 for $n \rightarrow \infty$,
\begin{eqnarray*}
&&\hat{s}_L \le M s, \quad  ||\hat{\bs \beta}_{L} -\bbeta_0||_1 \le M\sigma s \sqrt{{\log p \over n}},
\end{eqnarray*}
and
$$
||\mathbf{X}(\hat{\bs \beta}_{L} - \bbeta_0)||_2^2 \le M\sigma^2 s {\log p}.
$$
\end{lemma}

 \vspace{3ex}

{\bf A.3. Proof of Theorem~\ref{thm3}}

\vspace{2ex}
$(n-\hat{s}_L)(\hat{\sigma}^2_L -\sigma^2)$ can be decomposed as
\begin{eqnarray*}
&&(n-\hat{s}_L)(\hat{\sigma}^2_L -\sigma^2)\\
 &&= \big(\beps^T \beps - n \sigma^2\big) -
 2\cdot \beps^T \bX(\hat{\bs \beta}_{L}-\beta_0)   + ||\bX(\hat{\bs \beta}_{L}-\beta_0)||_2^2\\
&&=R_1 + R_2 + R_3.
\end{eqnarray*}
The classical central limit theorem yields $R_1 = O_P(n^{1/2})$. Note that
$$
|R_2| \le 2\cdot \big|\big|\bX^T \beps \big| \big |_\infty\cdot||\hat{\bs \beta}_{L}-\beta_0||_1.
$$
By (\ref{a16}) and Lemma 1, it follows that
$$
|R_2| =  O_P(\sqrt{n \log p}) \cdot O_P(s\sqrt{\log p/n}) = O_P(s \log p).
$$
In addition, by the third conclusion in Lemma 1, $|R_3|=O_P(s \log p)$. Therefore, the conclusion holds.

 \vspace{3ex}

{\bf A.4. Proof of Theorem~\ref{thm4}}

\vspace{2ex}
Let $\hat{\bbeta}^o = ({\hat{\bbeta}}_1^T, \mathbf{0}^T)^T$ with ${\hat{\bbeta}}_1 = ({\mathbf{X}}_1^{T} \mathbf{X}_1)^{-1}{\mathbf{X}}_1^{T} \mathbf{y}$ be the oracle estimator. The key step is to show that, with probability tending to 1, the oracle estimator $\hat{\bbeta}^o$ is a strictly local minimizer of $\bs{Q}_{n,\lambda_n}(\bbeta)$ defined by (\ref{a12}). To prove it, by Theorem 1 of  \cite{Fan:Lv:NP:2009}, it suffices to show that, with probability tending to 1, $\bs{\hat{\beta}}^o$ satisfies
\begin{eqnarray}
\label{apd-4-1}
&&\mathbf{X}_1^T(\mathbf{y} - {\mathbf{X}} \hat{\bbeta}^o) - n \tilde{\bs{\rho}}_{\lambda_n}(\hat{\bbeta}_1) = 0,\\
\label{apd-4-2}
&&||{\mathbf{X}}_2^T(\mathbf{y} - {\mathbf{X}} \hat{\bbeta}^o)||_{\infty} < n \rho'_{\lambda_n}(0+),\\
\label{apd-4-3}
&&\lambda_{\min}\big({1 \over n}\mathbf{X}_1^T \mathbf{X}_1\big) > \kappa_{\lambda_n}(\hat{\bbeta}_1),
\end{eqnarray}
where
$
\tilde{\bs{\rho}}_{\lambda_n}(\hat{\bbeta}_1) = \big(\mbox{sgn}(\hat{\beta}_1)\rho'_{\lambda_n}(|\hat{\beta}_1|),\cdots, \mbox{sgn}(\hat{\beta}_s)\rho'_{\lambda_n}(|\hat{\beta}_s|)\big)^T
$
and
$
\kappa_{\lambda_n}(\hat{\bbeta}_1) = \max_{j=1,\cdots, s}\big\{-\rho''_{\lambda_n}(|\hat{\beta}_j|)\big\}.
$

Let $\bs{\xi}_1 = \mathbf{X}_1^T \bs{\varepsilon}$ and $\bs{\xi}_2 = \mathbf{X}_2^T \bs{\varepsilon}$. Consider the events
$$
\mathcal{A}_{n1} = \Big\{||\bs{\xi}_1||_{\infty} \le \sqrt{n\log n \log\log n } \Big\} \cap
\Big\{\lambda_{\min}\big({1 \over n}\mathbf{X}_1^T \mathbf{X}_1\big) \ge c_1\Big\}
\hskip 0.5cm
\mbox{and}
$$
$$
%\hskip 0.5cm
\mathcal{A}_{n2} = \Big\{||\bs{\xi}_2||_{\infty} \le \sqrt{n^{\alpha_0 + 1}\log\log n}\Big\} \cap
\Big\{||{1 \over n}\mathbf{X}_2^T \mathbf{X}_1||_{2,\infty} \le c_2\Big\}. \hskip 1.5cm
$$
Observe that ${\hat{\bbeta}}_1 = ({\mathbf{X}}_1^{T} \mathbf{X}_1)^{-1}{\mathbf{X}}_1^{T} \mathbf{y}$. Then, we get
$
\hat{\bbeta}_1 - \bbeta_{01} = ({\mathbf{X}}_1^{T} \mathbf{X}_1)^{-1}{\mathbf{X}}_1^{T} \beps
$
and hence, under the event $\mathcal{A}_{n1}$,
\begin{eqnarray*}
||\hat{\bbeta}_1 - \bbeta_{01}||_\infty
&\le&||\hat{\bbeta}_1 - \bbeta_{01}||_2\\
&\le&||({1\over n}{\mathbf{X}}_1^{T} \mathbf{X}_1)^{-1}||_2||{1\over n}{\mathbf{X}}_1^{T} \beps||_2\\
&\le&\big[\lambda_{\min}({1\over n}{\mathbf{X}}_1^{T} \mathbf{X}_1)\big]^{-1}\cdot\sqrt{s}\cdot||{1 \over n}{\mathbf{X}}_1^{T} \beps||_\infty\\
&\le & c \sqrt{\log n \log\log n /n^{1-\alpha_0}} \ll \lambda_n,
\end{eqnarray*}
for some constant $c$ not depending on $n$. Note that, in the above inequalities, we use that facts $s = O(n^{\alpha_0})$ and $\lambda_n \propto n^{- {1-\alpha_0 \over 2}}\log n$.
%\begin{eqnarray*}
%||\hat{\bbeta}_1 - \bbeta_{01}||_\infty  \le ||\hat{\bbeta}_1 - \bbeta_{01}||_2  &\le & c \sqrt{\log n \log\log n /n^{1-\alpha_0}} \ll \lambda_n.
%\end{eqnarray*}

Since $d_n = {1 \over 2} \min_{j =1, \cdots,s} |\beta_{0j}| \ge n^{-\gamma} \log n $ with $\gamma \in (0, {1 / 2}]$ and $d_n \gg \lambda_n$, as addressed in Assumption (A9), we have, under the event $\mathcal{A}_{n1}$,
\begin{eqnarray*}
\min_{j= 1, \cdots,s}|\hat{\beta}_j| &\ge& \min_{j=1,\cdots,s}|\beta_{0j}| -  ||\hat{\bbeta}_1 - \bbeta_{01}||_\infty \\
 &\ge& 2\cdot d_n - c \sqrt{\log n \log\log n/n^{1-\alpha_0}} \\
 &\ge&  d_n \gg \lambda_n
\end{eqnarray*}
for sufficiently large $n$. As a result, this leads to
$
\tilde{\bs{\rho}}_{\lambda_n}(\hat{\bbeta}_1) = \bs{0}
$
and
$
\kappa_{\lambda_n}(\hat{\bbeta}_1) = 0,
$
and hence imply that (\ref{apd-4-1}) and (\ref{apd-4-3}) hold under the event $\mathcal{A}_{n1}$.

Now turn to prove the inequality (\ref{apd-4-2}).
%Note that $\lambda_n \propto n^{- {1-\alpha_0 \over 2}}\log n$ and $s = O(n^{\alpha_0})$.
Under the event $\mathcal{A}_{n1}\cap \mathcal{A}_{n2}$, we have
\begin{eqnarray} \label{a22}
||{1\over n}{\mathbf{X}}_2^T(\mathbf{y} - {\mathbf{X}} \hat{\bbeta}^o)||_{\infty}
&\le&  {1\over n}||\bs{\xi}_2||_{\infty} +
{1\over n }||{\mathbf{X}}_2^T {\mathbf{X}}_1||_{2,\infty}||\hat{\bbeta}_1 - \bbeta_{01}||_2\\ \nonumber
&\le& \sqrt{n^{\alpha_0-1}\log\log n} + c_2c\sqrt{\log n \log \log n/n^{1 - \alpha_0}} \\ \nonumber
& \propto& \lambda_n(\sqrt{\log\log n}/\log n + c_2c\sqrt{\log \log n/\log n}) \\ \nonumber
&\le& {1 \over 2} \lambda_n < \rho'_{\lambda_n}(0+)
\end{eqnarray}
for sufficiently large $n$.
This shows that the inequality (\ref{apd-4-2}) holds for sufficiently large $n$ under the event $\mathcal{A}_{n1} \cap \mathcal{A}_{n2}$. By taking $c=\sqrt{\log \log n}$, similar arguments to Theorem 1 lead to
$$
P\{\mathcal{A}_{n1} \cap \mathcal{A}_{n2}\} \rightarrow 1
$$
as $n \rightarrow \infty$. Thus, we have proven that $\hat{\bbeta}^o$ is a strictly local minimizer of $\bs{Q}_{n,\lambda_n}(\bbeta)$ with large probability tending to one. Consequently, $\hat \bbeta_{\sSCAD} = \hat{\bbeta}^o$.

 Now consider the asymptotic distribution of $\hat{\sigma}_{\sSCAD}^2 - \sigma^2$. Observe that $\hat{\bbeta}_1 = ({\mathbf{X}}_1^{T} \mathbf{X}_1)^{-1}{\mathbf{X}}_1^{T} \mathbf{y}$. Under the event $\mathcal{A}_{n1} \cap \mathcal{A}_{n2}$,
$$
\hat{\sigma}_{\sSCAD}^2- \sigma^2 = {1 \over n - s} \bs{\varepsilon}^T (\mathbf{I}_n - \textbf{P}_{M_0})\bs{\varepsilon} - \sigma^2.
$$
Hence, we have that
$$
\sqrt{n}\big(\hat{\sigma}_{\sSCAD}^2- \sigma^2 \big)\stackrel{\mathcal{D}}{\longrightarrow}  N(0, E[\varepsilon^4] - \sigma^4),
$$
which also implies that $\hat{\sigma}_{\sSCAD}^2 - \sigma^2  = O_P(n^{-1/2})$.
The proof is complete.

{\bf A.5. Proof of (\ref{a3a}) and (\ref{a3b})}

Let $\Phi(\cdot)$ and $F_{n-2}(\cdot)$ be the c.d.f. of standard Gaussian and student's $t$ distribution with $n-2$ degrees of freedom. For large $u$,
$$
1 - F_{n-2}(u) > 1 - \Phi(u) > \exp(-u^2).
$$
Therefore, $u=\sqrt{ \log (p/c)}$ satisfies $F_{n-2}(u)<\Phi(u)<1- c/p$.
The classical result that $\{\xi_{nj}\}_{j=1}^p$ are $i.i.d.$ $t_{n-2}$ distribution entails that
\begin{eqnarray*}
P\Big\{\sup_{1\leq j\leq p}\xi_{nj} > u \Big \}
&=&P\Big\{\sup_{1\leq j\leq p}F_{n-2}(\xi_{nj}) > F_{n-2}(u)\Big\}\\
&=& 1 - ( 1 - F_{n-2}(u) )^p,
\end{eqnarray*}
which, by the choice of $u$, is further bounded from below by
$$
  1- (1 - c/p)^p \geq 1 - \exp(-c).
$$

Note that $\gamma_{nj} = \xi_{nj} / (n-2 + \xi_{nj}^2)^{1/2}$ is strictly increasing.  It follows that
$$
  P\left \{ \sup_{1 \leq j \leq p} \gamma_{nj} > \frac{u}{(n-2 + u^2)^{1/2}} \right \}
  = P\Big\{\sup_{1\leq j\leq p}\xi_{nj} > u \Big \}
  \geq 1 - \exp(-c).
$$
The result (\ref{a3a}) follows from the fact that when $u^2 \leq n+2$,
$$
\frac{u}{(n-2 + u^2)^{1/2}} < \frac{u}{ \sqrt{2n}}.
$$

We now derive the limiting distribution (\ref{a3b}). For each $x > 0$,
\begin{eqnarray*}
    P\Big\{\sqrt{2\log p}\big(\sup_{1\leq j\leq p}\xi_{nj} - d_p\big) < x \Big\}
&=& P\Big\{ \sup_{1\leq j\leq p}\xi_{nj}   < d_p+ {x\over \sqrt{2\log p}}  \Big\} \\
&=& \Big(1-\int_{d_p+ {x\over \sqrt{2\log p}}}^{\infty}f_{n-2}(t)dt\Big)^p
\end{eqnarray*}
Therefore, it suffices to show
\begin{eqnarray}\label{a23}
p\int_{d_p+ {x\over \sqrt{2\log p}}}^{\infty}f_{n-2}(t)dt\rightarrow \exp\{-x\}.
\end{eqnarray}

Let $\nu = n-2$.  The following inequalities are helpful to verify the limit (\ref{a23})
\begin{eqnarray}\label{a24}
{1 \over \sqrt{2\pi}}\big({1 \over t} - {1 \over t^3}\big)\exp\big(-{t^2 \over 2}\big) \le \int_t^\infty \phi(s)ds \le \int_t^\infty f_{\nu}(s)ds \le
C(\nu){1 \over t}\cdot {\nu \over \nu -1}\Big(1  + {t^2 \over \nu}\Big)^{- {\nu -1 \over 2}},
\end{eqnarray}
where $C(\nu)={\Gamma({\nu+1\over 2})\over \sqrt{\nu \pi} \Gamma({\nu\over 2})}$.   Substituting $t=d_p+ {x\over \sqrt{2\log p}}$ into the inequalities (\ref{a24}), it is easy to verify that under condition $\log p=o(n^{\frac12})$,
$$\exp\{-x\}+o(1)<p\int_{d_p+ {x\over \sqrt{2\log p}}}^{\infty}f_{\nu}(t)dt< \exp\{-x\}+o(1).$$
This proves (\ref{a23}) and hence (\ref{a3b}).

%%%%%%%%%%%%%%%%%%%%%%%%%%%%%%%%%%%%%%%%%%%%%%%%%%%%%%%%%%%%%%%%%%%%%%%%%%%%%
%%%%%%%%%%%%%%%%%%%%%%%%%%%%%%%%%%%%%%%%%%%%%%%%%%%%%%%%%%%%%%%%%%%%%%%%
%\bibliographystyle{JASA_ref_harvard}
%\bibliography{ref_ve_JASA}

\begin{thebibliography}{}

\bibitem[\protect\citeauthoryear{Bickel, Ritov, and Tsybakov}{Bickel
  et~al.}{2009}]{bick:rito:simu:2008}
Bickel, P. J., Ritov, Y. and Tsybakov, A. (2009)
\newblock Simultaneous analysis of lasso and dantzig selector.
\newblock {\em Ann. Statist.,\/}~{\bf 37\/}, 1705--1732.

\bibitem[\protect\citeauthoryear{Bunea, Tsybakov, and Wegkamp}{Bunea
  et~al.}{2007}]{bune:tsyb:spar:2007}
Bunea, F., Tsybakov, A. and Wegkamp, M. (2007)
\newblock Sparsity oracle inequalities for the lasso.
\newblock {\em Elect. J. Statist.,\/}~{\bf 1}, 169--194.

\bibitem[\protect\citeauthoryear{Candes and Tao}{Candes and
  Tao}{2007}]{CandesTao07}
Candes, E. and Tao, T. (2007)
\newblock The dantzig selector: statistical estimation when $p$ is much larger than $n$ (with discussion).
\newblock {\em {Ann. Statist.,}\/}~{\bf {35}\/}, {2313--2351}.

\bibitem [Chatterjee and Lahiri(2010)]{ChatterjeeLahiri10}
Chatterjee A.  and Lahiri S. N. (2010) Bootstrapping lasso estimators. {\em Manuscript.}

\bibitem[\protect\citeauthoryear{Efron et al.}{Efron et al.}{2004}]{EHJT04}
Efron, B., Hastie, T., Johnstone, I. and Tibshirani, R. (2004)
\newblock Least angle regression (with discussions).
\newblock {\em {Ann. Statist.,}\/}~{\bf {32}\/}, {409-499}.


\bibitem[Fan and Li(2001)]{FanLi01} Fan, J. and Li, R. (2001). Variable selection via nonconcave
    penalized likelihood and its oracle properties.
    {\em Journal of American Statistical Association},
    {\bf 96}, 1348-1360.


\bibitem[\protect\citeauthoryear{Fan and Lv}{Fan and
  Lv}{2008}]{FanLv08}
Fan, J. and Lv, J. (2008)
\newblock Sure independence screening for ultrahigh dimensional feature space (with discussion).
\newblock {\em J. R. Statist. Soc. \/}~{\bf B, 70\/}, 849--911.

\bibitem[\protect\citeauthoryear{Fan and Lv}{Fan and Lv}{2009}]{Fan:Lv:NP:2009}
Fan, J. and Lv, J. (2009)
\newblock Properties of non-concave penalized likelihood with  NP-dimensionality.
\newblock {\em Manuscript\/}.

\bibitem[\protect\citeauthoryear{Fan and Lv}{Fan and
  Lv}{2010}]{FanLv10}
Fan, J. and Lv, J. (2010)
\newblock A selective overview of variable selection in high dimensional
  feature space.
\newblock {\em Statist. Sinica.,\/}~{\bf 20}, 101--148.

\bibitem[\protect\citeauthoryear{Fan and Peng}{Fan and
  Peng}{2004}]{ISI:000221981400004}
Fan, J. and Peng, H. (2004)
\newblock {Nonconcave penalized likelihood with a diverging number of parameters}.
\newblock {\em {Ann. Statist.,}\/}~{\bf {32}\/}, {928--961}.

\bibitem[\protect\citeauthoryear{Fan, Samworth, and Wu}{Fan
  et~al.}{2009}]{Fan:Samworth:Wu:2008}
Fan, J., Samworth, R. and Wu, Y. (2009)
\newblock Ultrahigh dimensional feature selection: Beyond the linear model.
\newblock {\em J. Mach. Learn., Res.,\/}~{\bf 10}, 2013--2038.

\bibitem[\protect\citeauthoryear{Fan and Song}{Fan and
  Song}{2010}]{FanSong2010}
Fan, J. and Song, R. (2010)
\newblock Sure independence screening in generalized linear models with NP-dimensionality.
\newblock {\em Ann. Statist.\/}, to appear.

\bibitem[\protect\citeauthoryear{Greenshtein and Ritov}{Greenshtein and
  Ritov}{2004}]{Greenshtein:Ritov:persistence:2004}
Greenshtein, E. and Ritov, Y. (2004)
\newblock Persistence in high-dimensional linear predictor selection and the virtue of overparametrization.
\newblock {\em Bernoulli,\/}~{\bf 10\/}, 971--988.

\bibitem[Han {\em et al.}(2010)]{HanGuFan10}  Han, X., Gu, W., and Fan, J. (2010). Control of the false
discovery rate under arbitrary covariance dependence.  {\em Manuscript.}


\bibitem[\protect\citeauthoryear{Kim, Choi, and Oh}{Kim
  et~al.}{2008}]{Kim:Choi:OH:SCADHD:2008}
Kim, Y., Choi, H. and Oh, H.-S. (2008)
\newblock Smoothly clipped absolute deviation on high dimensions.
\newblock {\em J. Am. Statist. Assoc.,\/}~{\bf 103},
  1665--1673.

\bibitem [Knight and Fu(2000)]{KnightFu00}
Knight, K. and Fu, W.(2000) Asymptotics for lasso-type estimators.
     {\em Ann. Statist.}, {\bf 28}, 1356-1378.

\bibitem [Kyung, {\it et al.}(2010)]{MGMC2010}
     Kyung M., Gill, J., Ghosh M. and Casella G. (2010) Penalized regression, standard errors and Bayesian lassos.
{\em Bayesian analysis}, {\bf 5(2)}, 369-412.

\bibitem[Lv and Fan(2009)]{LvFan09}
Lv, J. and Fan, Y. (2009). A unified approach to model selection and sparse recovery using regularized least squares. \textit{Ann. Statist.} \textbf{37}, 3498--3528.

\bibitem [Meier et al. (2008)]{MeierGB08}
        Meier, L., van de Geer, S. and B\"uhlmann, P. (2008). The group
        LASSO for logistic regression.
        {\em Journal of the Royal Statistical Society, B}, 70, 53-71.

\bibitem [Meinshausen, {\it et al.}(2009)]{MMB09}
Meinshausen, N., Meier, L. and B$\ddot{u}$hlmann P. (2009) p-Values for high-dimensional regression. {\em J. Am. Statist. Assoc.,} {\bf 104}, 1671-1681.

\bibitem[\protect\citeauthoryear{Meinshausen and Yu}{Meinshausen and
  Yu}{2009}]{mein:yu:lass:2009}
  Meinshausen, N. and Yu, B. (2009)
\newblock LASSO-type recovery of sparse representations for high-dimensional data.
\newblock {\em Ann. Statist.,\/}~{\bf 37\/}, 246--270.

\bibitem [Park and Casella(2008)]{ParkCasella08}
Park, T. and Casella, G. (2008) The Bayesian lasso. {\em J. Am. Statist. Assoc.,} {\bf 103}, 681-686.

\bibitem[\protect\citeauthoryear{Wasserman and Roeder}{Wasserman and
  Roeder}{2009}]{Wasserman:Roeder:HDVS:2009}
Wasserman, L. and Roeder, K. (2009)
\newblock High dimensional variable selection.
\newblock {\em Ann. Statist.,\/}~{\bf 37\/}, 2178--2201.



\bibitem[van der Vaart and  Wellner(1996)]{VW96}
         van der Vaart, A.W. and  Wellner, J.A. (1996).  Weak Convergence and
         Empirical Processes.  Springer, New York.

\bibitem[\protect\citeauthoryear{Ye}{Ye}{1998}]{Ye:on:1998}
Ye, J. (1998)
\newblock On measuring and correcting the effects of data mining and model  selection.
\newblock {\em J. Am. Statist. Assoc.,\/}~{\bf 93},
  120--131.

\bibitem[\protect\citeauthoryear{Zhang and Huang}{Zhang and
  Huang}{2008}]{ISI:000258243000006}
Zhang, C. H. and Huang, J. (2008)
\newblock The sparsity and bias of the lasso selection in high-dimensional  linear regression.
\newblock {\em Ann. Statist.,\/}~{\bf 36\/}, {1567--1594}.


\bibitem[\protect\citeauthoryear{Zhao and Li}{Zhao and
  Li}{2010}]{ZhaoLi2010}
Zhao, S. and Li, Y. (2010)
\newblock {Principled sure independence screening for Cox models with ultra-high-dimensional covariates}.
\newblock {Preprint}~{}{}.

\bibitem[\protect\citeauthoryear{Zhao and Yu}{Zhao and
  Yu}{2006}]{ISI:000245390700010}
Zhao, P. and Yu, B. (2006)
\newblock {On model selection consistency of lasso}.
\newblock {\em J. Mach. Learn. Res.,\/}~{\bf {7}},
  {2541--2563}.

\bibitem[\protect\citeauthoryear{Zou}{Zou}{2006}]{Zou:adap:2006}
Zou, H. (2006)
\newblock The adaptive Lasso and its oracle properties.
\newblock {\em J. Am. Statist. Assoc.,\/}~{\bf 101\/}, 1418--1429.

\bibitem [Zou and Li(2008)]{ZouLi08}
     Zou, H. and Li, R. (2008). One-step Sparse Estimates in Nonconcave Penalized Likelihood Models (with discussion).  {\em Ann. Statist.}, {\bf 36}, 1509-1533.

\bibitem [Zou, {\it et al.}(2007)]{ZHT07}
Zou, H., Hastie, T. and Tibshirani, R (2007). On the ``degrees of freedom'' of the lasso. {\em Ann. Statist.}, {\bf 35}, 2173-2192.
\end{thebibliography}
%\bibliographystyle{Chicago}

\end{document}